\documentclass[acmsmall]{acmart}

\usepackage{graphicx}
\usepackage{amsfonts}
\usepackage{subcaption}
\usepackage{caption}
\usepackage[title]{appendix}
\usepackage{colortbl}
\usepackage{pifont}
\usepackage{multirow}
\usepackage{enumitem} 
\usepackage{moreenum}
\usepackage{breakcites}
\usepackage{threeparttable}
\usepackage[normalem]{ulem} 
\usepackage{arydshln}
\usepackage{amsthm}
\usepackage{amsmath}
\usepackage{mathtools}

\usepackage{lscape} 
\usepackage[normalem]{ulem}
\newcommand{\cmark}{\ding{51}}%
\usepackage{todonotes}
\usepackage{booktabs}
\usepackage{tabularx}
\usepackage{array}
\newcolumntype{L}[1]{>{\raggedright\let\newline\\\arraybackslash\hspace{0pt}}m{#1}}
\newcolumntype{C}[1]{>{\centering\let\newline\\\arraybackslash\hspace{0pt}}m{#1}}
\newcolumntype{R}[1]{>{\raggedleft\let\newline\\\arraybackslash\hspace{0pt}}m{#1}}
\definecolor{brightpink}{rgb}{1.0, 0.0, 0.5}
\definecolor{blue(pigment)}{rgb}{0.2, 0.2, 0.7}
\definecolor{White}{gray}{0.995}
\theoremstyle{plain}
\newtheorem{thm}{Theorem} 

\theoremstyle{definition}
\newtheorem{exmp}[thm]{Example} 
\DeclareMathOperator{\sign}{sign}

\usepackage{tikz}
\usepackage{pgfplots}

\AtBeginDocument{%
  \providecommand\BibTeX{{%
    \normalfont B\kern-0.5em{\scshape i\kern-0.25em b}\kern-0.8em\TeX}}}

\acmJournal{CSUR}	
\acmVolume{1}
\acmNumber{1}
\acmArticle{1}
\acmYear{2020}
\acmMonth{4}

\usepackage{pifont}
\usepackage{soul}
\DeclareMathOperator*{\argmax}{arg\,max}

\definecolor{White}{gray}{0.995}
\usepackage{colortbl}%
  \newcommand{\myrowcolour}{\rowcolor[gray]{0.925}}
   
\newcommand{\dquotes}[1]{``#1''}

\usepackage{mathtools}

\definecolor{blue(pigment)}{rgb}{0.2, 0.2, 0.7}

\def\updated#1{{\color{black}#1}}

\usepackage{tabularx}

\usepackage{makecell}

\usepackage{parskip}
\usepackage{appendix}

\renewcommand{\qed}{\hfill\rule{1ex}{1ex}}

\makeatletter
\renewcommand\subsubsection{\@startsection{subsubsection}{3}{\z@}%
                       {-18\p@ \@plus -4\p@ \@minus -4\p@}%
                       {4\p@ \@plus 2\p@ \@minus 2\p@}%
                       {\normalfont\normalsize\bfseries\boldmath\itshape
                        \rightskip=\z@ \@plus 8em\pretolerance=10000 }}
\makeatother

\begin{document}

\title[A survey on Adversarial Recommender Systems: from Attack/Defense strategies to Generative Adversarial Networks]{A survey on Adversarial Recommender Systems: from Attack/Defense strategies to Generative Adversarial Networks}

\author{Yashar Deldjoo}
\orcid{0000-0002-6767-358X}
\email{yashar.deldjoo@poliba.it}
\author{Tommaso Di Noia}
\orcid{0000-0002-0939-5462}
\email{tommaso.dinoia@poliba.it}
\author{Felice Antonio Merra}
\authornote{Authors are listed in alphabetical order. Corresponding author: Felice Antonio Merra.}
\email{felice.merra@poliba.it}
\orcid{1234-5678-9012}
\affiliation{%
  \institution{Polytechnic University of Bari}
  \streetaddress{Via Orabona, 4}
  \city{Bari}
  \state{Italy}
  \postcode{70125}
}


\renewcommand{\shortauthors}{Y. Deldjoo, T. Di Noia and F.A. Merra}

\begin{abstract}
Latent-factor models (LFM) based on collaborative filtering (CF), such as matrix factorization (MF) and deep CF methods, are widely used in modern recommender systems (RS) due to their excellent performance and recommendation accuracy.  However, success has been accompanied with a major new arising challenge: \textit{many applications of machine learning (ML) are adversarial in nature}~\cite{DBLP:series/synthesis/2018Vorobeychik}. In recent years, it has been shown that these methods are vulnerable to adversarial examples, i.e., subtle but non-random perturbations designed to force recommendation models to produce erroneous outputs. 



The goal of this survey is two-fold: (i) to present recent advances on adversarial machine learning (AML) for the security of RS (i.e., attacking and defense recommendation models), (ii) to show another successful application of AML in generative adversarial networks (GANs) for generative applications, \updated{thanks to their ability for learning (high-dimensional) data distributions.} In this survey, we provide an exhaustive literature review of \updated{74} articles published in major RS and ML journals and conferences. This review serves as a reference for the RS community, working on the security of RS or on generative models using GANs to improve their quality.



\end{abstract}





\maketitle

\section{Introduction}\label{sec:Introduction}
In the age of data deluge, where users face a new form of information explosion, recommender systems (RS) have emerged as a paradigm of information push to lessen decision anxieties and consumer confusion by over-choice. RS enhance users' decision-making process and support sales
by personalizing item recommendations for each user and helping them discover novel products. RS are a pervasive part of user experience online today and serve as the primary choice for many consumer-oriented companies such as Amazon, Netflix, and Google (e.g., YouTube~\cite{DBLP:conf/imc/ZhouKG10}).
Among different recommendation techniques, collaborative filtering (CF) methods have been the mainstream of recommendation research both in academia and industry due to their superb recommendation quality. CF builds on the fundamental assumption that users who have expressed similar interests in the past will maintain similar choices in future~\cite{DBLP:journals/cacm/GoldbergNOT92}, and infers target user preference over unseen items by leveraging behavioral data of other users and exploiting similarities in their behavioral patterns.


\begin{figure}
    \centering
    \includegraphics[width = 0.90\linewidth]{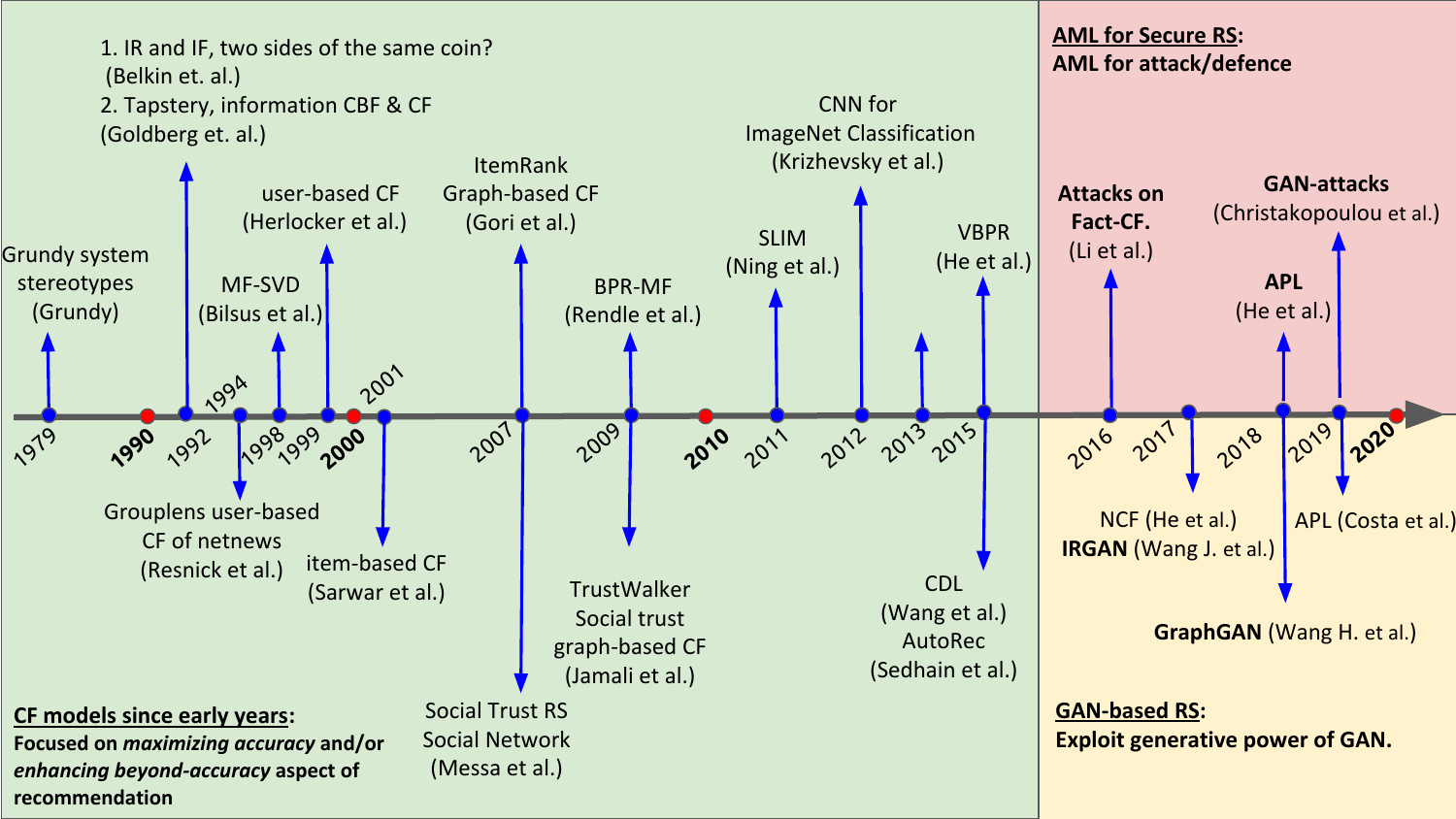}
    \caption{Milestones of CF recommender models.}
    \label{fig:milestones}
\end{figure}


Milestones in CF models over the last three decades are illustrated in Figure~\ref{fig:milestones}. We can identify two major eras in development of CF models based on their main objective:


\begin{enumerate}
    \item The era focused on maximizing/enhancing the recommendation accuracy and beyond-accuracy; 
    \item The post neural era, the transition era from classical learning to adversarial machine learning.
\end{enumerate}

\noindent \textbf{Accuracy maximization and beyond-accuracy enhancement era.} In this era, the main effort of research and practitioner-scholars was concentrated on the \dquotes{golden objective} of \textit{maximizing recommendation accuracy}. Consequently, machine-learned models tend to use any available signal in the data to reach this goal, even though some of the data contained noise as the results of users' misoperations. We distinguish between two classes of CF techniques in this era: (i) classical non-neural CF, (ii) deep neural CF, each described in the following.

\begin{itemize} 
    \item \textbf{Classical non-neural CF.} The starting of this era dates back to the 1990s and is still progressing. Over these three decades, the study on CF methods has been the subject of active research by the RS community resulting in a diverse set of models and evaluation measures to assess the effectiveness of these models. We can classify these CF approaches based on various dimensions. For example, from a \textit{learning paradigm} perspective, CF models can be classified according to (i) \textit{memory-based CF} and (ii) \textit{model-based CF} models, in which the former category makes recommendation based on the similarity of users-user interactions (i.e., user-based neighborhood model) or item-item interactions (i.e., item-based neighborhood model) while the latter category predicts users' feedback of unseen items using latent factor models such as matrix-factorization (MF)~\cite{DBLP:journals/computer/KorenBV09}. From the \textit{model training} perspective, it is possible to categorize these models based on the loss functions employed according to (i) \textit{point-wise} loss where the goal is to optimize towards a predefined ground-truth (e.g., matrix factorization approach based on SVD), (ii) \textit{pairwise ranking loss} where the goal is to optimize personalized ranking  (e.g., matrix factorization based on BPR) and (iii) \textit{list-wise} loss where the objective is to reflect the distance between the reference list and the output list ~\cite{DBLP:conf/recsys/ShiLH10}.

	\item \textbf{Deep neural CF.}  Another milestone is concerned with the success of  \dquotes{neural} technology in machine learning (ML). DNNs have shown to be capable of providing remarkable accuracy in several predictive tasks and domains such as image classification~\cite{DBLP:journals/spm/HanZCLX18} and speech recognition~\cite{DBLP:journals/ijar/SalakhutdinovH09} among others. In the field of RS, DNNs have been shown useful for the recommendation in several ways such as extracting deep features (via using CNNs), modeling item content in CF models by integrating side item information, building CF models by parameterizing latent factor models into layers of a DNN (deep CF), and modeling sequential relations (via using RNNs). As for deep-CF approaches, while MF assumes that the \textit{linear interaction} between user and item latent factors can explain observed feedback, deep CF models can model a more complex representation of hidden latent factors by \textit{parametrization of MF via a DNN}.   
\end{itemize}

\updated{The above system types have been redesigned to use a wealth of side information beyond the URM into the recommendation models to make RS adapted in specific domains. The surveys~\cite{DBLP:journals/csur/ShiLH14, deldjoo2020,deldjoo2020multimedia} provide a good frame of reference for CF methods leveraging rich side information.} 






\noindent \textbf{The post neural era, the transition era from classical learning to adversarial machine learning.} Despite the significant success of DNNs to solve a variety of complex prediction tasks on non-structured data such as images, recently, they have been demonstrated to be vulnerable to \textit{adversarial examples}. Adversarial examples (or adversarial samples) are subtle but non-random perturbations \textit{designed} to dictate a ML model to produce erroneous outputs (e.g., to misclassify an input sample). The subject started booming after the pioneering work by~\citet{DBLP:journals/corr/SzegedyZSBEGF13} reported the vulnerability of DNNs against adversarial samples for the image classification task. It has been shown that by adding a negligible amount of adversarial perturbation on an image (e.g., a panda), a CNN classifier could misclassify the image in another class (e.g., a gibbon) with high confidence.  These results were quite shocking since it was expected that state-of-the-art DNNs that generalize well on unknown data do not change the label of a test image that is slightly perturbed and is human-imperceptible. Algorithms that aim to find such adversarial perturbations are referred to as \textit{adversarial attacks}. As ML models are involved in many consumer safety and security-intensive tasks such as autonomous driving, facial recognition, and camera surveillance, adversarial attacks pose significant concerns to the security and integrity of the deployed ML-models.

In the field of RS, numerous works have reported the failure of machine-learned recommendation models, i.e., latent-factor models (LFM) based on CF like MF and deep CF methods widely adopted in modern RS, against adversarial attacks. For instance, \citet{DBLP:conf/sigir/0001HDC18}  showed that by exposing
the model parameters of BPR~\cite{DBLP:conf/uai/RendleFGS09} to both adversarial and random perturbations of the BPR model parameters, the value of nDCG is decreased by -21.2\% and -1.6\% respectively, which is equal to a staggering impact of approximately 13 times difference. One main explanation for such behavior is that adversarial attacks exploit the imperfections and approximations made by the ML model during the training phase to control the models' outcomes in an engineered way~\cite{DBLP:journals/corr/PapernotMG16}.

Adversarial machine learning (AML) is an emerging research field that combines the best practices in the areas of ML, robust statistics, and computer security~\cite{DBLP:conf/ccs/HuangJNRT11, DBLP:journals/ijon/XiaoBNXER15}. It is concerned with the design of learning algorithms that can resist adversarial attacks, studies the capabilities and limitations of the attacker, and investigates suitable countermeasures to design more secure learning algorithms~\cite{DBLP:conf/ccs/HuangJNRT11}.
The pivotal distinguishing characteristic of AML is the notion of \dquotes{min-max} game, in which two competing players play a zero-sum differential game, one --- i.e., the attacker --- tries to \textit{maximize} the likelihood of the attack success, while the other --- i.e., the defender --- attempts to \textit{minimize} the risk in such a worst-case scenario. In the context of RS, the defender players can be a machine-learned model such as BPR or a neural network, while the attacker is the adversarial model.

To protect models against adversarial attacks, \textit{adversarial training} has been proposed. It is a defensive mechanism whose goal is not to detect adversarial examples, instead to build models that perform equally well with adversarial and clean samples. Adversarial training consists of injecting adversarial samples ---generated via a specific attack model such as FGSM~\cite{DBLP:journals/corr/GoodfellowSS14} or BIM~\cite{DBLP:conf/iclr/KurakinGB17}--- into each step of the training process. It has been reported ---both in RS~\cite{8618394} and ML~\cite{DBLP:journals/corr/abs-1911-05268}--- that this process leads to robustness against adversarial samples (based on the specific attack type on which the model was trained on), and better generalization performance on clean samples. For instance, in~\cite{8618394}, the authors show that the negative impact of adversarial attacks measured in terms of nDCG is reduced from -8.7\% to -1.4\% when using adversarial training instead of classical training.

The above discussion highlights the failure of classical ML models (trained on clean data) in adversarial settings and advocates the importance of AML as a new paradigm of learning to design more secure models. Nevertheless, the attractiveness of AML that exploits the power of two adversaries within a \dquotes{min-max} game is not limited to security applications and has been exploited to build novel \textit{generative} models, namely generative adversarial networks (GANs). The key difference is as follows: the models used in AML for security (or attack and defense) focus only on a class of discriminative models (e.g., classifiers), whereas GANs build upon both discriminative and generative models.
A GAN is composed of two components: the generator $G$ and the discriminator $D$. The training procedure of a GAN is a min-max game between $G$,  optimized to craft fake samples such that $D$ cannot distinguish them from real ones, and $D$, optimized to classify original samples from generated ones correctly. 
Through the interplay between these two components, the model reaches the Nash equilibrium where $G$ has learned to mimic the ground-truth data distribution, e.g., a profile of a particular user. In the present survey, we identified different application for GAN-based RS that include, improving negative sampling step in learning-to-rank objective function~\cite{DBLP:conf/kdd/WangYHLWH18, DBLP:conf/ijcai/FanD0WTL19}, fitting the generator to predict missing ratings by leveraging both temporal~\cite{DBLP:conf/ijcai/ZhaoWYGYC18, DBLP:conf/recsys/BharadhwajPL18} and side-information~\cite{DBLP:conf/sigir/WangYZGXWZZ17, DBLP:conf/cikm/ChaeKKL18}, or augmenting training dataset~\cite{DBLP:conf/www/ChaeKKC19, DBLP:journals/tmm/DuFYXCT19}.

\subsection{Main contributions and outcome of the survey}

\updated{The adversarial learning paradigm uses a unified and theoretically principled \dquotes{min-max} optimization formulation at its heart to advance research in adversarial robustness. Nonetheless, the utility of this general min-max optimization has been demonstrated in another category of applications named GANs, to learn (high-dimensional) data distributions. The current survey at hands aims to help the reader to learn fundamental concepts in these two  application areas of AML and provide a comprehensive literature review on the new trend of AML research in the RecSys field.}

\begin{enumerate}
    \item \textit{AML for the robustness/security of RS}: This is the \dquotes{principal application} of AML in RS, which focuses on adversarial attacks on and defense of recommendation models, which we present it in Section~\ref{sec:security}.
    \item \textit{Application of AML in GANs}: This is a \dquotes{derived topic} from AML that is focused on \dquotes{generative} learning models. We identified four types of applications in this category, namely: \textit{improving CF recommendation}, \textit{context-aware recommendation}, \textit{cross-domain recommendation} and \textit{complementary recommendation}, which we present in Section~\ref{sec:GAN}.
\end{enumerate}

Overall, AML-based recommendation scenarios are highly relevant to the field of RS. Indeed, in recent years, a growing number of relevant research works have been proposed. Despite this success, research in AML-RS is overly scattered with each paper focusing on a particular task, domain, or architecture.
One major objective of this survey is to categorize state-of-the-art research in the field based on several identified dimensions in order to provide a richer understanding of the different facets of the AML-RS. 
Our ultimate motivation is to lay the foundation for a more standardized approach for reproducible research works in the field.
 
The practical outcome of the present survey includes:
\begin{enumerate}
    \item To the best of our knowledge, this is the first work that provides a \textit{comprehensive understanding} of AML in RS domain, unifying the advances made in the communities of ML and RS;
    \item This survey sheds lights on two successful applications of AML, namely: adversarial attacks and defenses and GANs, both using the concept of \dquotes{min-max} game at their core. It provides an extensive literature review of the existing research, specifically:
    \begin{itemize}
        \item For AML-based RS focused on security: we present a unified problem formulation and discuss the existing adversarial attack studies on RS from various perspectives in particular attack and defense models, recommendation dimensions as well as evaluation and attack metrics used in different papers.
        \item For GAN-based RS, we provide a conceptual view of recommendation approaches incorporating GAN to address the item recommendation task and we review an extensive number of research, which we classify according the generator, discriminator type and training paradigm. We also categorize the existing research into several distinctive high-level goals (e.g., complementary recommendation in fashion domain, context-aware recommendation, etc.).
     \end{itemize}
    \item We created an open-source repository\footnote{Table with AML-RS publications at \url{https://github.com/sisinflab/adversarial-recommender-systems-survey}} that includes all reviewed research articles which is updated over time. The aim of this repository is to facilitate bench-marking AML in the RS field by proving the released codes links and datasets used for the evaluation.
\end{enumerate}


\subsection{Strategies for literature search}
To identify the relevant publications that constitute the state-of-the-art on adversarial learning in recommender systems, we mainly relied on publications indexed in major computer science bibliography databases namely DBLP (\url{https://dblp.uni-trier.de/}) and Scopus (\url{https://www.scopus.com}). In addition, realizing the fact that many top-tiers venues also publish related works, which may not be necessarily indexed in the above databases, we also gathered a number of related publications by searching directly through Google Scholar. Our search strategy was composed of two main stages: 
\begin{enumerate}
\item relevant publication collection, 
\item filtering and preparing the final list.
\end{enumerate} 
We collected also referenced publications in the yet selected ones.
As for the first stage, we queried the term \dquotes{adversarial recommend} in the above-mentioned indexing services. While search in DBLP returns publications containing the query term in the \textit{title}, the search results from Scopus and Google Scholar return publications containing the query \textit{both in the tile and the content}, thereby all-together forming a complete list of identified research works. We collected all resulting publications from DBLP, Scopus and Google Scholar search. In the second stage, we went through all the collected research works and removed all irrelevant works. These for instance could include works that used AML for an application different than RS e.g., in Computer Vision, and Speech Enhancement.
We mostly turned our attention to conference-level and journal publications published in top conferences and to a lesser extent to workshop publications or works published in entry-level venues. Some of the considered journals and conferences include: \updated{RecSys, SIGIR, WSDM, TheWebConference, IJCAI, and KDD}. 


\subsection{Survey Context and Related Surveys}

\updated{While there exist several survey articles on general RS topics, for example~\citet{DBLP:journals/csur/ShiLH14,deldjoo2020,DBLP:journals/csur/QuadranaCJ18, DBLP:journals/fthci/EkstrandRK11,DBLP:journals/kbs/BobadillaOHG13},
to the best of our knowledge, none of them focuses on the application of AML techniques in the recommendation task. In contrast, we provide a comprehensive literature review and extended taxonomy on the application of AML for security purposes (i.e., adversarial attacks and defenses) and generative models (i.e., in GAN-based systems). This classification is further accompanied with the identification of main application trends, possible future directions, and novel open challenges.}

\updated{The current literature review can be seen nonetheless related to other surveys such as~\cite{DBLP:journals/csur/ZhangYST19, DBLP:journals/corr/abs-2003-05730}.  In particular,~\citet{DBLP:journals/csur/ZhangYST19} introduce AML as a novel application of DL by pointing out to very few works of the field such as~\cite{DBLP:conf/sigir/WangYZGXWZZ17, DBLP:conf/sigir/0001HDC18} without providing a detailed study on the topic of AML for RS. The work by~\citet{DBLP:journals/corr/abs-2003-05730}, on the other hand, is centered around the application of AML for graph learning in general ML setting. Although, link prediction techniques can be adapted from graph-learning based system to perform item recommendation task (e.g., by predicting a user's connection with an item), this work remains far from the focus of the current survey. We would like also to acknowledge existence of related surveys on the application of AML in other tasks, not directly related to RS, for example for the CV field by~\citet{akhtar2018threat}, on classical ML models by~\citet{DBLP:journals/access/LiuLZCYL18}, and GAN applications in CV and NLP tasks by~\citet{DBLP:journals/access/PanYYKYZ19}.}

We can also highlight that part of the material presented in this survey has been presented as a tutorial at the WSDM'20~\cite{DBLP:conf/wsdm/DeldjooNM20} and the RecSys'20\cite{DBLP:conf/recsys/AnelliDNM20} conference.\footnote{Tutorial slides at ~\url{https://github.com/sisinflab/amlrecsys-tutorial}}

\subsection{Structure of the Survey}
\updated{In the subsequent core section of this survey, we
present foundation concepts to AML and its application for security of ML and RS models in Section~\ref{sec:security}. Afterward, in Section~\ref{subsec:AML_RS}, we present a literature review on state-of-the-art research on the application of AML for the security of recommendation models and categorize the research works based on several identified dimensions. Section~\ref{sec:GAN} sheds light on another exciting application for AML in GAN-based systems and reveals several applications of that in the RecSys domain. Section~\ref{sec:conclusion} rounds off the survey by presenting the findings and open research directions.}





\begin{table}[t]
\centering
\caption{List of abbreviations used throughout this paper.} 
\label{tbl:abbr}
\scalebox{0.80}{

 \begin{tabular}{l|l}
\toprule

\multicolumn{1}{c}{\textbf{Abbreviation}} & \multicolumn{1}{c}{\textbf{Term}} \\

\bottomrule
AI & Artificial Intelligence \\ \hline
AML & Adversarial Machine Learning \\ \hline
AMR & Adversarial Multimedia Recommendation \\ \hline
C\&W &Carlini and Wagner \\\hline
CA & Context-Aware RS \\ \hline
CBF-RS & Content-Based Filtering RS\\ \hline
CF-RS & Collaborative Filtering RS \\ \hline
CS & Cold start \\ \hline
CD-RS & Cross-Domain RS\\ \hline
CV & Computer Vision \\ \hline
DL & Deep Learning \\ \hline
DNN & Deep Neural Network \\ \hline
ERM & Empirical Risk Minimization \\ \hline 
FGSM & Fast Gradient Sign Method  \\ 
\bottomrule
\end{tabular}

\quad

 \begin{tabular}{l|l}
\toprule

\multicolumn{1}{c}{\textbf{Abbreviation}} & \multicolumn{1}{c}{\textbf{Term}} \\

\bottomrule

FNCF & Feature-based Neural Collaborative Filtering \\ \hline
GAN & Generative Adversarial Network \\ \hline
G-RS & Graph-based RS \\ \hline
GRU & Gated Recurrent Unit \\ \hline
IR & Information Retrieval \\ \hline
LFM & Latent Factor Model \\ \hline
LSTM & Long Short-Term Memory \\ \hline
MF & Matrix Factorization \\ \hline
ML & Machine Learning \\ \hline
NLP & Natural Language Processing \\ \hline
RS & Recommender Systems \\ \hline
SM & Social Media \\ \hline
SN & Social Network \\ \hline
URM & User Rating Matrix \\ 

\bottomrule
\end{tabular}

}
\end{table}

\section{Adversarial Machine Learning for Security of RS}\label{sec:security}
For security concerns to be addressed appropriately in today's ML systems, there is a need to bridge the knowledge gap between the ML and computer security communities. Adversarial machine learning (AML) is a recently proposed and popularized approach that lies at the intersection of the above fields combining the best works of the two. The main goal of AML for security is to build systems that can learn in normal conditions and such that when they are under attack ---in particular under \textit{adversarial attack}--- they can respond rapidly and
safeguard ML models against emerging adversaries' threats.

As the literature on AML for security emerged in the context of ML, in Section~\ref{sec:security}, we first discuss the fundamentals of  attacks on, and defenses of ML models (cf. Section~\ref{subsec:AML_att_def}). We then present AML-based RS focused on security applications in which we survey various identified literature in the field and classify them based on several methodological and evaluation-related dimensions (cf. Section~\ref{subsec:AML_RS}).

\subsection{Foundations of adversarial machine learning}
\label{subsec:AML_att_def}
Throughout this section, we consider a supervising learning --- classification --- task. Assume a training dataset $\mathcal{D}$ of $n$ pairs $(x, y) \in \mathcal{X} \times \mathcal{Y}$, where $x$ is the input sample, and $y$ is its corresponding class label. The problem of classification is formulated as finding a candidate function $f_\theta: \mathcal{X} \rightarrow \mathcal{Y}$  that can predict the class label $y$ around the input sample $x$, where $\theta$ is the model parameter.  This leads to solving an empirical risk minimization (ERM) problem of the form 
\[
\min_{\theta} \sum_{(x_i,y_i) \in \mathcal{D}} \ell(f(x_i;\theta),y_i)
\] 
where $\ell(.)$ is the empirical risk function (the loss function). Various adversarial attacks aim to find a non-random perturbation $\delta$ to produce an adversarial example $x^{adv} = x + \delta$ that can cause an erroneous prediction (e.g., misclassification) as we will see in the following section.
\subsubsection{\updated{Adversarial attacks on ML models}}
In recent years, the advances made in deep learning (DL) have considerably advanced the intelligence of ML models in a unique number of predictive tasks such as classification of images and other unstructured data. Notwithstanding their great success, recent studies have shown that ML/DL models are not immune to security threats from adversarial use of AI. We can classify attacks against a ML model along three main dimensions, attack \textit{timing} and \textit{goal}.
    \begin{figure}[!t]
    \centering
    \includegraphics[width=0.70\paperwidth]{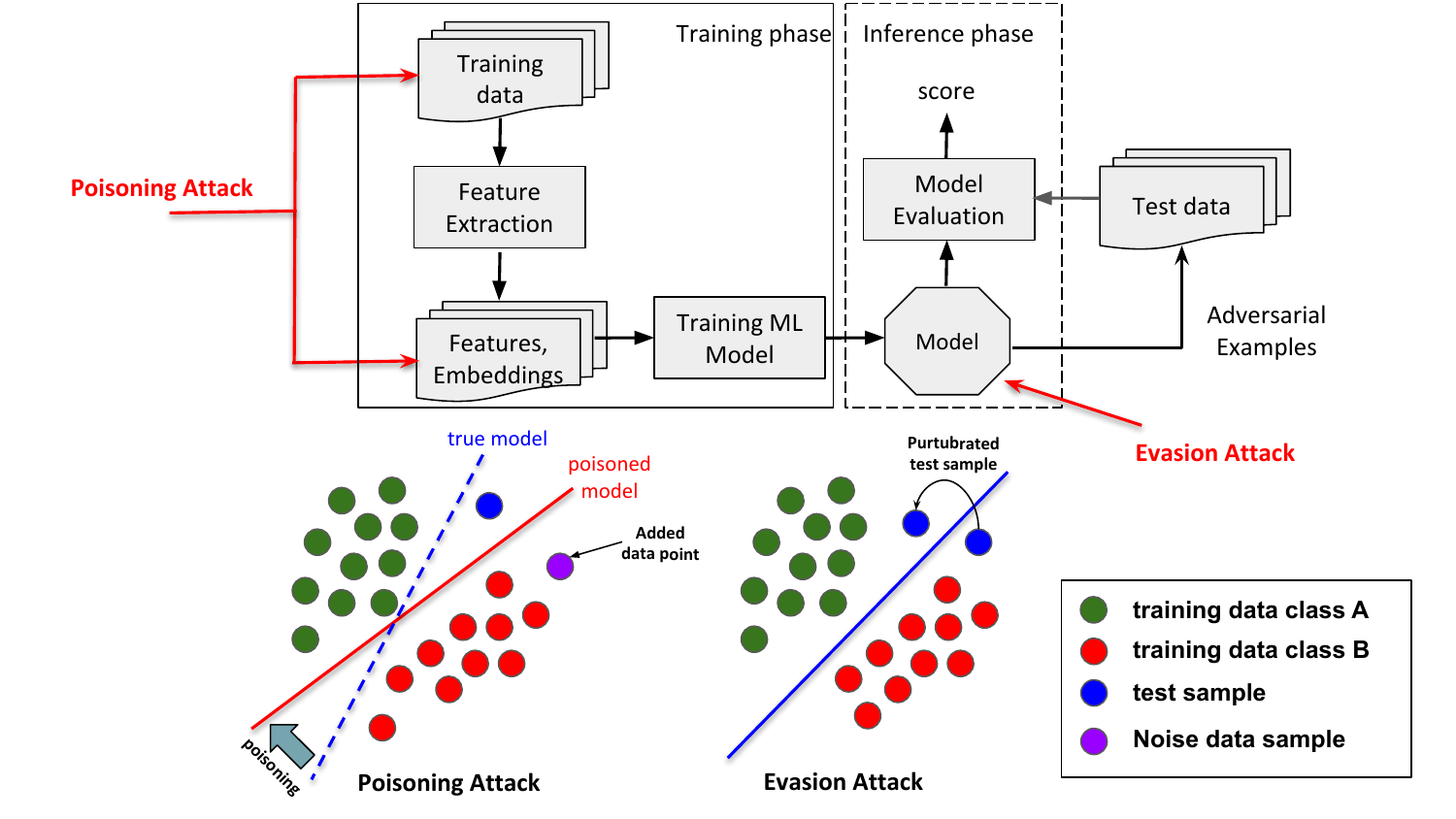}
    \caption{A schematic representation of the distinction between \textit{evasion} attacks and \textit{poisoning} attacks.}
    \label{figure:attack_scheme}
\end{figure}

\textbf{Attack timing.} As illustrated in Fig.~\ref{figure:attack_scheme}, an adversary can attack a ML model at two main stages of the learning pipeline, during \textit{training} or \textit{production}. These two categories of attacks are respectively known as (i) \textit{training-time attack} (a.k.a. causative or poisoning attack)~\cite{DBLP:conf/icml/BiggioNL12} and, ii) \textit{inference-time attack} (a.k.a. exploratory or evasion attack)~\cite{DBLP:journals/corr/SzegedyZSBEGF13}.
    \begin{itemize}
    \item \textit{Poisoning attack.} Data poisoning attacks are realized by injecting false data points into the training data with the goal to corrupt/degrade the model (e.g., the classifier). Poisoning attacks have been explored in the literature for a variety of tasks~\cite{DBLP:series/synthesis/2018Vorobeychik}, such as (i) attacks on binary classification for tasks such as label flipping or against kernelized SVM~\cite{DBLP:conf/ecai/XiaoXE12}, (ii) attacks on unsupervised learning such as clustering and anomaly detection~\cite{DBLP:journals/corr/abs-1811-09985}, and (iii) attacks on matrix completion task in RS~\cite{DBLP:conf/nips/LiWSV16, DBLP:conf/recsys/DeldjooNM19}. As an example, in the pioneering work by~\citet{DBLP:conf/icml/BiggioNL12}, the authors propose a poisoning attack based on properties of the SVM optimal solution that could significantly degrade the classification test accuracy. 
    \item \textit{Evasion attack.} \updated{Unlike poisoning attacks, evasion attacks do not interfere with training data. They adjust malicious samples during the inference phase. These attacks are also named \textit{decision-time} attacks referring to their attempt to \textit{evade} the \textit{decision} made by the learned model at test time~\cite{DBLP:series/synthesis/2018Vorobeychik}. For instance, evasion attacks can be used to evade spam~\cite{DBLP:journals/jmlr/JorgensenZI08}, as well as network intrusion~\cite{DBLP:conf/raid/WangPS06} detectors. Recently, evasive attacks are conducted by crafting \textit{adversarial examples}, which are subtle but non-random human-imperceptible perturbations, added to original data to cause the learned model to produce erroneous output.~\citet{DBLP:journals/corr/SzegedyZSBEGF13} were the first to discover that some carefully selected perturbations that are barely perceptible to the human eye, when added to an image, could lead a well-trained DNN to misclassify the adversarial image with high confidence.}

    
    \end{itemize}
    
\textbf{Attack goal.} Attacks are conducted for different goals. We can distinguish between two main classes of attack goals: i) \textit{untargeted attack} and, ii) \textit{targeted attack}. To provide the reader with an intuitive insight of the mechanism behind adversarial attacks and defense strategies, we define them formally for a classification task~\cite{DBLP:series/synthesis/2018Vorobeychik}.


\begin{definition}[Untargeted adversarial attack~\cite{DBLP:journals/corr/SzegedyZSBEGF13}]\label{def:untar_attack} The goal of the attacker in \textbf{untargeted adversarial attack} (misclassification) is to add a minimal amount of perturbation $\delta$ on the input sample $x$ such that it can cause  incorrect classification.
 Given $f(x;\theta) = y$, an Untargeted Adversarial Attack is formulated as:
    \begin{equation}
        \label{eq:att_untrg}
        \begin{aligned}
            & \min_{\delta}
            & & \left\lVert \delta \right\rVert \\
            & \text{s.t.:}
            & & f(x + \delta; \theta) \neq y , \ \ x + \delta \in [0,1]^n
        \end{aligned}
    \end{equation}
The second constraint $x + \delta \in [0,1]^n$ is a value-clipping constraint needed for images, to bound the adversarial samples into to a predefined range so that the images remain visible after adversarial attack. Alternatively, we can formulate the problem as an unconstrained optimization problem where the goal of the attacker is to \textit{maximize} the loss between the perturbed sample $x + \delta$ and true class $y$
    \begin{equation}
        \label{eq:att_rel_gen}
        \begin{aligned}
            & \underset{\delta: \left\lVert \delta \right\rVert \leq \epsilon }{\text{max}}
            & &\ell(f(x+\delta; \theta),y) 
        \end{aligned}
    \end{equation}
Obviously since adding an unbounded amount of noise on the input will eventually lead to a classification error, the goal of the attacker is to minimize a norm-constrained form of noise, that is $\left\lVert \delta \right\rVert \leq \epsilon$ for some exogenously given $\delta$. \qed
\end{definition}
In the context of DNN, the above attacks are categorized based on the norm used to represent the magnitude of the noise according to the following norm types~\cite{DBLP:series/synthesis/2018Vorobeychik}: $l_0$, $l_1$ and $l_2$ and $l_{\infty}$.

\begin{definition}[Targeted adversarial attack~\cite{DBLP:journals/corr/SzegedyZSBEGF13}]
    \label{def:untar_attack}
The goal of the attacker in \textbf{targeted adversarial attack} is to perturb the input by adding a minimum amount of perturbation $\delta$ such that it can force the model to misclassify the perturbed sample into an illegitimate target class (aka mis-classification label). Given $f(x;\theta) = y$, with $y \neq y_t$, we formulate the problem as:
    \begin{equation}
        \label{eq:att_trg}
        \begin{aligned}
            & \underset{\delta}{\text{min}}
            & & \left\lVert \delta \right\rVert \\
            & \text{s.t.:}
            & & f(x + \delta; \theta) = y_t
        \end{aligned}
    \end{equation}
Similarly, the above problem can be expressed as a unconstrained optimization problem 
    \begin{equation}
        \label{eq:att_rel_gen}
        \begin{aligned}
            & \underset{\delta: \left\lVert \delta \right\rVert \leq \epsilon }{\text{min}}
            & & \ell(f(x+\delta; \theta), y_t) 
        \end{aligned}
    \end{equation}  
    \qed 
\end{definition}
The most common attack types so far exploited in the community of RS are fast gradient sign attack (FGSM)~\cite{DBLP:journals/corr/GoodfellowSS14} and Carlini and Wagner (C\&W) attacks, which belong to $l_{\infty}$- and $l_2$-norm   attack types respectively. We provide the formal definition of the FGSM and C\&W attacks here.
\begin{definition}[FGSM attack~\cite{DBLP:journals/corr/GoodfellowSS14}]
The fast gradient sign method (FGSM)~\cite{DBLP:journals/corr/GoodfellowSS14} utilizes the \textit{sign of the gradient} of the loss function to find perturbation that maximizes the training loss (for untargeted case)
    \begin{equation}
        \label{eq:att_linf}
        \delta = \epsilon \cdot \sign    (\bigtriangledown_x \ell(f(x;\theta),y))
    \end{equation}
where $\epsilon$ (perturbation level) represents the attack strength and $\bigtriangledown_x$ is the gradient of the loss function w.r.t. input sample $x$. The adversarial example is generated as $x^{adv} = x + \delta$. FGSM applies an $l_{\infty}$-bound constraint $||\delta||_{\infty} \leq \epsilon$ with the original idea to encourage perceptual similarity between the original and perturbed samples. The unconstrained FGSM  aims to find perturbation that would increase/maximize the loss value. The corresponding approach for targeted FSGM~\cite{DBLP:conf/iclr/KurakinGB17} is
    \begin{equation}
        \label{eq:att_linf_targeted}
        \delta = - \epsilon \cdot \sign    (\bigtriangledown_x \ell(f(x;\theta),y_t))
    \end{equation}
where the goal is to maximize the conditional probability $p(y_t|x)$ for a given input $x$.\qed 
\end{definition}

Several variants of the FGSM has been proposed in the literature~\cite{DBLP:journals/corr/abs-1911-05268,DBLP:journals/corr/abs-1810-00069}. For instance, the fast gradient value (FGV) method~\cite{DBLP:conf/cvpr/RozsaRB16}, which instead of using the sign of the gradient vector in FGSM, uses the actual value of the gradient vector to modify the adversarial change, or basic iterative method (BIM)~\cite{DBLP:conf/iclr/KurakinGB17}  (a.k.a iterative FGSM) that applies FGSM attack multiple times \textit{iteratively} using a small step size and within a total acceptable input perturbation level.

\begin{definition}[C\&W attack~\cite{DBLP:conf/sp/Carlini017}]
The Carlini and Wagner (C\&W) attack~\cite{DBLP:conf/sp/Carlini017} is one of the most effective attack models. The core idea of C\&W attack is to replace the standard loss function --- e.g., typically cross-entropy --- with an empirically-chosen loss function and use it in an \textit{unconstrained optimization formulation} given by
\begin{equation}
    \underset{\delta}{\text{min}} \left\lVert \delta \right\rVert_p^p + c \cdot h(x+\delta, y_t)
\end{equation}
where $h(\cdot)$ is the candidate loss function.\qed 
\end{definition}
The C\&W attack has been used with several norm-type constraints on perturbation $l_0$, $l_2$, $l_{\infty}$ among which the $l_2$-bound constraint has been reported to be most effective~\cite{DBLP:journals/corr/CarliniW16}.


\textbf{Adversarial attacks on RS - challenges and differences with ML tasks.} In spite of the similarities between ML classification and recommendation learning tasks, there are considerable differences/challenges in adversarial attacks on RS compared with ML and the degree to which the subject has been studied in the respective communities:
\begin{itemize}
    \item \textit{Poisoning vs. adversarial attack.} In the beginning, the main focus of RS research community has been on \textit{hand-engineered} fake user profiles (a.k.a shilling attacks) against rating-based CF~\cite{deldjoo2020dataset}. Given a URM with $n$ real users and $m$ items, the goal of a shilling attack is to augment a fraction of malicious users $\lfloor{\alpha n} \rfloor$ ($\lfloor{.}\rfloor$ is the floor operation) to the URM ($\alpha \ll 1$) in which each malicious use profile can contain ratings to a maximum number of $C$ items. The ultimate goal is to harvest recommendation outcomes toward an illegitimate benefit, e.g., pushing some targeted items into the top-$K$ list of users for market penetration. Shilling attacks against RS have an established literature and their development face two main milestones: the first one ---since the early 2000s--- where the literature was focused on building hand-crafted fake profiles whose rating assignment follow different strategy according to random, popular, love-hate, bandwagon attacks among others~\cite{DBLP:journals/air/GunesKBP14}; the second research direction started in 2016 when the first ML-optimized attack was proposed by~\citet{DBLP:conf/nips/LiWSV16} on factorization-based RS. This work reviews a novel type of data poisoning attack that applies the adversarial learning paradigm for generating poisoning input data. Nonetheless, given their significant impact against modern recommendation models, the research works focusing on \textit{machine-learned adversarial attacks} against RS have recently received great attention from the research community, \updated{e.g., consider~\cite{DBLP:conf/sigir/CaoCY0Z20,  DBLP:conf/sigir/LiuXC0YZ20}.}
    
    \item \textit{CF vs. classification models:} Attacks against classification tasks focus on enforcing the wrong prediction of individual instances in the data. In RS, however, the mainstream attacks rely on CF principles, i.e., mining similarity in opinions of like-minded users to compute recommendations. This interdependence between users and items can, on the one hand, \textit{improve robustness} of CF, since predictions depend on a group of instances not on an individual one and, on the other other hand, may cause \textit{cascade effects}, where attack on individual user may impact other neighbor users~\cite{DBLP:conf/recsys/Christakopoulou19}.
   
    \item \textit{Attack granularity and application type:} Adversarial examples created for image classification tasks are empowered based on continuous real-valued representation of image data (i.e., pixel values), but in RS, the raw values are user/item IDs and ratings that are discrete. Perturbing these discrete entities is infeasible since it may lead to changing the semantics of the input, e.g., loosely speaking applying $ID+\delta$ can result in a new user $ID$. Therefore, existing adversarial attacks in the field of ML are not transferable to the RS problems trivially. Furthermore, in the context of CV --- attacks against images --- the perturbations often need to be \dquotes{human-imperceptible} or \dquotes{inconspicuous} (i.e., may be visible but not suspicious)~\cite{DBLP:journals/corr/abs-1911-05268}. How can we capture these nuances for designing attacks in RS remains as an open challenge.
\end{itemize}

\subsubsection{Defense against adversarial attacks}
From a broad perspective, defense mechanisms against adversarial attacks can be classified as \textit{detection} methods and methods seeking to increase the \textit{robustness} of the learning model. The goal of this section is to briefly review approaches that build \textit{robust ML} models in adversarial settings. The prominent methods used in RS are (i) the robust optimization~\cite{DBLP:conf/iclr/KurakinGB17, DBLP:conf/sigir/0001HDC18} and, (ii) the distillation method~\cite{DBLP:conf/sp/PapernotM0JS16, DBLP:journals/tois/ChenZXQZ19}.

\textbf{Robust optimization against adversarial attacks.}
At the heart of the robust optimization method is the assumption that every sample in the training data $\mathcal{D}$ can be a source for adversarial behavior. It performs an ERM against a specific adversary on each sample in $\mathcal{D}$ and applies a zero sum-game between the prediction and attack adversaries
leading to the following robust optimization framework
\begin{equation}
    \begin{aligned}
        \label{eq:robust_opt_2}
         \min_{\theta}  \sum_{(x_i,y_i) \in \mathcal{D}} \max_{\delta: \left\lVert \delta \right\rVert \leq \epsilon} \ell(f(x_i+\delta; \theta),y_i)
    \end{aligned}
\end{equation}
where $\epsilon$ is an upper-bound on the adversarial perturbation level $\delta$. The ultimate goal in robust optimization is that the prediction model will perform equally well with adversarial and clean inputs.

\begin{definition}[Adversarial training~\cite{DBLP:journals/corr/GoodfellowSS14}]
The goal of adversarial training is to build a robust model from ground-up on a training set augmented with adversarial examples. Adversarial regularization is one of the mostly investigated techniques for adversarial training, which utilizes an approximation of the worst-case loss function, i.e., $\max_{\delta: \left\lVert \delta \right\rVert \leq \epsilon} \ell(f(x + \delta; \theta),y_i)$, as the regularizer.
\begin{equation}
    \label{eq:robust_opt_adv_reg}
        \ell_{T} = \underbrace{\min_{\theta}  \sum_{i \in \mathcal{D}} [\ell(f(x; \theta),y_i) + \lambda \underbrace{\max_{\delta: \left\lVert \delta \right\rVert \leq \epsilon} \ell(f(x + \delta; \theta),y_i)}_\text{optimal attack model}}_\text{optimal robustness-preserving prediction}]
\end{equation}
\qed 
\end{definition}
As it can be noted, the inner maximization finds the strongest attack  against the prediction model that is subject to adversarial perturbation. The outer minimization estimates the strongest defensive against a given attack by giving up a level of accuracy due to the regularization. The parameter $0<\lambda<1$ controls the trade-off between accuracy (on clean data) and robustness (on perturbed data).

\begin{exmp}[Adversarial training of BPR-MF]
BPR is the state-of-the-art method for personalized ranking  implicit feedbacks. The main idea behind BPR is to maximize the distance between positively and negatively rated items. Given the training dataset $D$ composed by positive and negative items for each user, and the triple $(u,i,j)$ (user $u$, a positive item $i$ and negative item $j$), the BPR objective function is defined as 
\begin{equation}
    \label{eq:BPR_loss}
    \ell_{BPR}(\mathcal{D} | \Theta) = \argmax_{\Theta} \sum_{(u,i,j) \in \mathcal{D}} ln \, \sigma(\hat{x}_{ui}(\Theta)-\hat{x}_{uj}(\Theta))-\lambda\left \| \Theta \right \|^2
\end{equation}
where $\sigma$ is the logistic function, and $\hat{x}_{ui}$ is the predicted score for user $u$ on item $i$ and $\hat{x}_{uj}$ is the predicted score for user $u$ on item $j$; $\lambda \left \| \Theta \right \|^2$ is a regularization method to prevent over-fitting.\footnote{As it can be noted, BPR can be viewed as a classifier on the triple $(u,i,j)$, where the goal of the learner is to classify the difference $\hat{x}_{ui}-\hat{x}_{uj}$ as correct label +1 for a positive triple sample and 0 for a negative instance.} Adversarial training of BPR-MF similar to Eq.~\ref{eq:robust_opt_adv_reg} can be formulated as
\begin{equation}
    \label{eq:robust_bpr_adv_reg}
    \ell_{APR} = \underbrace{\min_{\theta}  \sum_{(u,i,j) \in D} [\ell_{BPR}(\mathcal{D} | \Theta) + \lambda \underbrace{\max_{\delta: \left\lVert \delta \right\rVert \leq \epsilon} \ell_{BPR}(\mathcal{D} | \Theta + \delta)]}_\text{optimal attack model against BPR}}_\text{optimal robustness preserving defensive}
\end{equation} \qed 
\end{exmp}

\begin{figure}[!t]
    \centering
    \includegraphics[width=0.7\paperwidth]{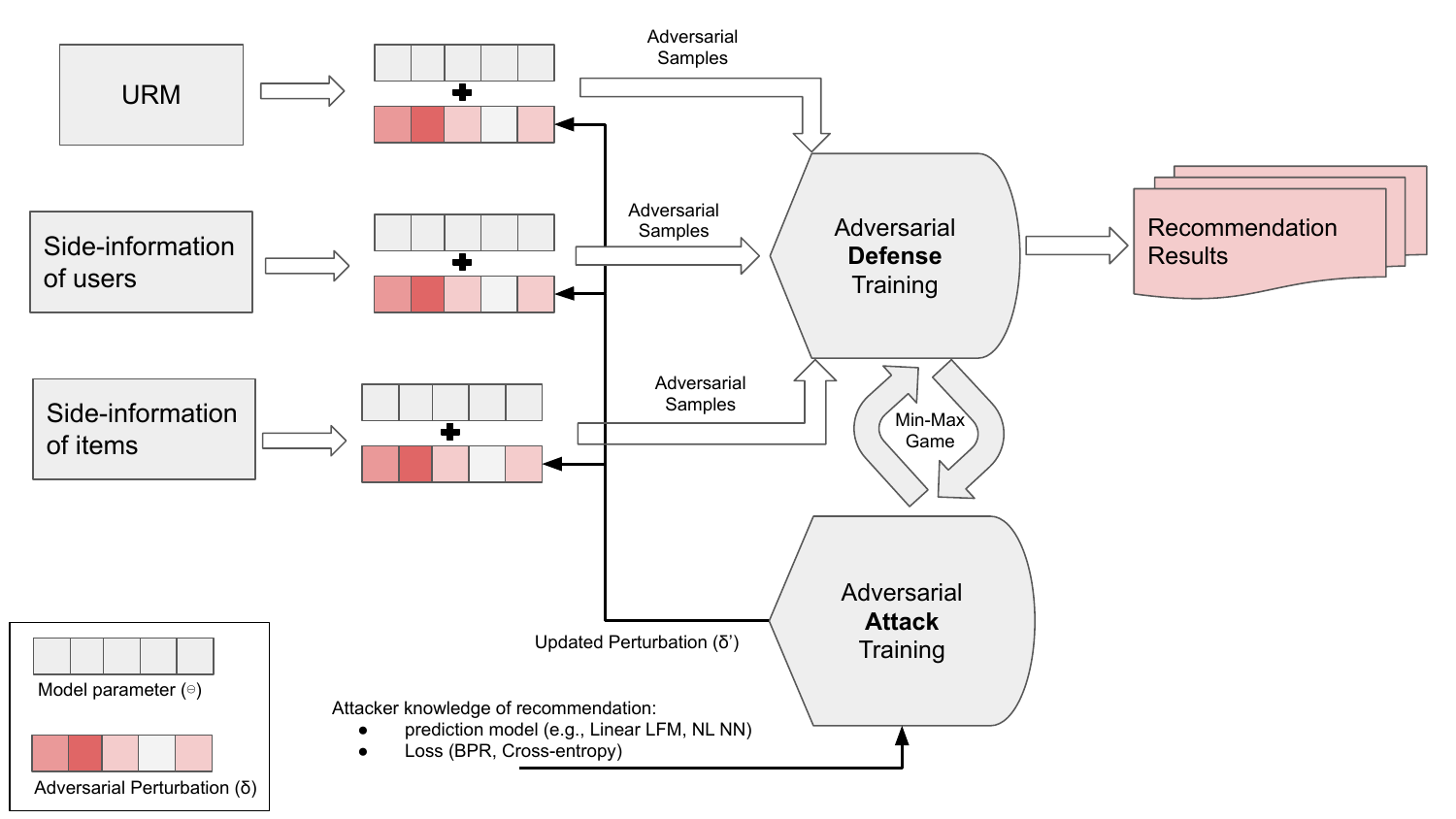}
    \caption{A notional view of Adversarial Recommendation Framework integrating the adversarial perturbations on users and items, and their side information, model parameters.}
    \label{figure:adv_framework}
\end{figure}
We do not report details on \textit{distillation} \cite{DBLP:journals/corr/HintonVD15} as defense strategy since it is not very common for RS.

\subsection{Adversarial Machine Learning for Attack and Defense on RS}\label{subsec:AML_RS}
In this section, we focus on state-of-the-art approaches to the application of AML in RS research. RS which employ AML for security applications in recommendation tasks, follow the simplified steps sketched in Fig.~\ref{figure:adv_framework}.
In the following, in addition to providing concise summaries of the surveyed works, for a convenient overview, we categorize the reviewed research articles in Table~\ref{tbl:adl_attack} according to the following dimensions:


\begin{itemize}
     \item \textbf{Model.} This column lists the model name and provides the reference to the main paper.
     
     \item \textbf{Attack and Defense Model.} This column represents the main \textit{attack} and \textit{defense} strategies applied on various recommendation models and the \textit{attack granularity} on the system.
     \begin{enumerate}
         \item \textit{Attack model.}
         Among all attacks strategies proposed in the community of CV~\cite{DBLP:journals/access/AkhtarM18}, in RS the most dominant attack approaches to date have been \textit{FGSM} and \textit{C\&W}, \updated{and \textit{Other} strategies (e.g., multi-step adversarial attacks~\cite{DBLP:conf/iclr/KurakinGB17a, DBLP:conf/iclr/MadryMSTV18}, GAN-based attack models~\cite{DBLP:conf/recsys/Christakopoulou19,DBLP:conf/cikm/LinC0XLY20}))}
         \item \textit{Defense model.}  As for the best defensive strategy against attack, we have found the strategy \textit{adversarial training (a.k.a. adversarial regularization)} as the most commonly-adopted approach irrespective of the attack model, while  \textit{distillation} is adopted only by a single paper~\cite{DBLP:journals/tmm/DuFYXCT19}.
         \item \textit{Attack granularity.} This column represents the level of data on which the adversarial perturbation is added on. It is important to note that while in the computer vision domain, these perturbations are added on raw data (e.g., pixel values), in RS, they are applied on the model parameters of recommendation strategy, as illustrated in Fig.~\ref{figure:adv_framework}. In particular, adversarial perturbations are added to one of the following data: (i) directly on the \textit{user profile }(i.e., user rating profile), (ii) \textit{user and item latent factor model parameters} in an LFM, e.g., according to $\mathbf{p}'_u = \mathbf{p}_u + \delta$, $\mathbf{q}'_i = \mathbf{q}_i + \delta$ in which $\mathbf{p}_u, \mathbf{q}_i \in \mathbb{R}^F$ are $F$-dimensional embedding factors  whose linear interaction explains an unobserved preference; (iii) and (iv) \textit{embeddings representing side information of user and items} respectively.
    \end{enumerate}
     
     \item \textbf{Recommendation \& Learning.}
     The core recommendation models that we consider in this survey are CBF, CF and CA. We also consider hybrid systems but we do not specify a placeholder for such systems; if an approach use both CBF+CF, we simply mark both corresponding columns, regardless of which hybridization technique it uses~\cite{aggarwal2016ensemble}. Instead, given the ML (optimization)-based approach for most of the considered papers, we categorize papers based on the recommendation prediction model according to \textit{linear LFM} (e.g., MF or variations of that such as PMF), \textit{linear tensor factorization} (TF), \textit{non-linear models based on auto-encoder} (NL-AE) and \textit{neural network} (NL-NN); furthermore we classify the loss function used in the core optimization model of the attack and defense scenarios based on \textit{BPR}~\cite{DBLP:conf/uai/RendleFGS09} and \textit{cross-entropy}.

\end{itemize}

\updated{To further help categorization of the research works, we split them according to three adversarial objectives, according to adversarial attacks and defense against: (i) accuracy of recommendations, (ii) privacy of users and (iii) bias and fairness of recommendation models as shown in Table~\ref{tbl:adl_attack}.}


\begin{table}[!t]
\caption{Classification of approaches that address \textit{adversarial learning for attacking and defending} RS models} \label{tbl:adl_attack}
{\scalebox{0.70}{
\begin{tabular}{p{1.90cm} c c|c c c c c c c c c|c c c c c c c c c}
\toprule
\multicolumn{1}{c}{\textbf{Name}}
&\multicolumn{1}{c}{\textbf{Authors}}
&\multicolumn{1}{c}{\textbf{Year}} & \multicolumn{9}{c}{\textbf{Attack \& Defense Models}} & \multicolumn{9}{c}{\textbf{Recommendation \& Learning}} \\ 
\cmidrule(r){1-1}\cmidrule(r){2-2}\cmidrule(r){3-3}\cmidrule(lr){4-12}\cmidrule(lr){13-21}
&
&
& \multicolumn{3}{c}{\begin{tabular}[c]{@{}c@{}}attack \\ model\end{tabular}} 
& \multicolumn{2}{c}{\begin{tabular}[c]{@{}c@{}}defense \\ model\end{tabular}} 
& \multicolumn{4}{c|}{\begin{tabular}[c]{@{}c@{}}attack \\ granularity \end{tabular}}
& \multicolumn{3}{c}{\begin{tabular}[c]{@{}c@{}}Rec class\end{tabular}} 
& \multicolumn{4}{c}{\begin{tabular}[c]{@{}c@{}}predictor\\ model type\end{tabular}} 
& \multicolumn{2}{c}{\begin{tabular}[c]{@{}c@{}} loss \\function\end{tabular}}\\
\cmidrule(lr){4-6}\cmidrule(lr){7-8}\cmidrule(lr){9-12}\cmidrule(lr){13-15}\cmidrule(lr){16-19}\cmidrule(lr){20-21}
&
&
&\rotatebox{90}{\small{FGSM}} &\rotatebox{90}{\small{C\&W}} &\rotatebox{90}{\small{Others}}  &\rotatebox{90}{\small{Adv. Training}} &\rotatebox{90}{\small{Distillation}}  &\rotatebox{90}{\small{user profile}}  &\rotatebox{90}{\small{user embed}} &\rotatebox{90}{\small{item embed}} &\rotatebox{90}{\small{\begin{tabular}[c]{@{}c@{}}side embed\end{tabular}}}  &\rotatebox{90}{\small{CBF}}  &\rotatebox{90}{\small{CF}}  &\rotatebox{90}{\small{CA}}  &\rotatebox{90}{\small{Linear LFM}}  &\rotatebox{90}{\small{Linear TF}}  &\rotatebox{90}{\small{NL AE}}  &\rotatebox{90}{\small{NL NN}} 
&\rotatebox{90}{\small{BPR}}   &\rotatebox{90}{\small{Cross-entropy}}  
\\
\bottomrule
\multicolumn{3}{l|}{\textbf{Accuracy}} &  & & & &   &  &  &  & & &  &   &  & & & &  &   \\ \hline 
\small{APR~\cite{DBLP:conf/sigir/0001HDC18}} &He at al. &2018 &\cmark &  &  & \cmark &  &  &\cmark  &\cmark  &  & &\cmark    &  &\cmark  & & & &\cmark   &  \\ [1ex]
\myrowcolour
\small{AMR~\cite{8618394}} &Tang et al. &2019 &\cmark &  &  & \cmark &   &  &  &  & \cmark &\cmark &\cmark    &  &\cmark & & & & \cmark  &   \\ [2ex]
\small{FGACAE~\cite{DBLP:conf/sigir/YuanYB19}} &Yuan et al. &2019 &\cmark &  &  & \cmark & &  &  \cmark &  \cmark &  & &\cmark    &  & & &\cmark & &  &\cmark    \\ [1ex]
\myrowcolour
\small{ACAE~\cite{DBLP:conf/ijcnn/YuanYB19}} &Yuan et al. &2019 &\cmark &  & & \cmark &  &  & \cmark &  \cmark &  & &\cmark &  & & &\cmark & &  &\cmark  \\ [1ex]
\small{FNCF~\cite{DBLP:journals/tmm/DuFYXCT19}} &Du et al. &2019 & &\cmark  &  & & \cmark &  & \cmark & \cmark &  & &\cmark &  & & & &\cmark &  &  \\ [1ex]
\myrowcolour
\small{ATF~\cite{DBLP:conf/recsys/ChenL19}} &Chen et al. &2019 &\cmark &  &  &\cmark &  &  &\cmark  & \cmark &\cmark  & &\cmark &\cmark  & &\cmark & & &\cmark  &   \\ [1ex]
\small{GANAtt~\cite{DBLP:conf/recsys/Christakopoulou19}} &\small{Christakopoulou} et al. &2019 & & & \cmark   & &   & \cmark &  &  &  & & \cmark &  & \cmark& & & &  &  \\ [1ex]
\myrowcolour
\small{AdvIR~\cite{DBLP:conf/www/ParkC19}} &Park et al. &2019 & \cmark & & & \cmark&   & &\cmark  & \cmark &  & &  \cmark &  & \cmark & & & & \cmark  & \\ [1ex]
\small{AMASR~\cite{DBLP:conf/sigir/TranSL19}} &Tran et al. &2019 & \cmark &  & &\cmark &  &  & \cmark & \cmark &  & &\cmark &  & & & & \cmark & \cmark  & \\ [1ex]
\myrowcolour
\small{ATMBPR~\cite{DBLP:journals/access/WangH20a}} &Wang et al. &2020 & \cmark &  & & \cmark & &  & \cmark & \cmark &  & & \cmark &  &  \cmark & & & & \cmark  & \\  [1ex]
\small{SACRA~\cite{DBLP:conf/wsdm/LiW020}} &Li et al. &2020 &\cmark &  &  & \cmark & & &\cmark  &\cmark  &  &   &\cmark & \cmark & \cmark  & & & \cmark  &\cmark   & \\ [1ex]
\myrowcolour 
\small{MSAP~\cite{anelli2020multi}} & Anelli et al. & 2020 & \cmark &  &  \cmark &  \cmark &  &  & \cmark & \cmark &   & \cmark  &  &  & \cmark & & & & \cmark  &  \\ [1ex]
\small{TAaMR~\cite{DBLP:conf/dsn/DiNoia20}} & Di Noia et al. & 2020 & \cmark & & & &   &  &  &  & \cmark  & \cmark  &  \cmark &   & \cmark & & & & \cmark  &  \\ [1ex]
\myrowcolour 
\small{AIP~\cite{DBLP:journals/corr/abs-2006-01888}} & Liu et al. & 2020 & & & \cmark & &   &  &  &  & \cmark  & \cmark  &  \cmark &   & \cmark & & & & \cmark  &  \\ [1ex]
\small{SAO~\cite{DBLP:conf/sigir/ManotumruksaY20}} & \small{Manotumruksa} et al. & 2020 & \cmark & &  & \cmark  &   &  & \cmark & \cmark &  &  &  & \cmark &  & & & \cmark & \cmark  &  \\ [1ex]
\myrowcolour 
\small{VAR~\cite{anelli2020empirical}} & Anelli et al. & 2020 & \cmark &  \cmark &  \cmark &  \cmark &  &  &  &  & \cmark  & \cmark  &  \cmark &   & \cmark & \cmark & & & \cmark  &  \\ [1ex] 
\small{AUSH~\cite{DBLP:conf/cikm/LinC0XLY20}} & Lin et al. & 2020 & & & \cmark   & &   & \cmark &  &  &  & & \cmark &  & \cmark & & \cmark & \cmark &  &  \\ [1ex]
\hline

\multicolumn{3}{l|}{\textbf{Privacy}} &  & & & &   &  &  &  & & &  &   &  & & & &  &   \\ \myrowcolour
\hline
\small{PAT~\cite{DBLP:conf/icpram/ResheffESS19}} & Resheff et al. & 2019 & &  &  & \cmark &   &  &\cmark  &  &  &  & \cmark &   & \cmark & & & & \cmark  &  \\ [1ex]
\small{RAP~\cite{DBLP:conf/wsdm/BeigiMGAN020}} & Beigi et al. & 2020 & &  & \cmark  & \cmark &   &  &\cmark  &  &  &  & \cmark &   & \cmark & & & & \cmark  &  \\ [1ex]
\hline
\multicolumn{3}{l|}{\textbf{Bias, Fairness}} &  & & & &   &  &  &  & & &  &   &  & & & &  &   \\
\hline
\myrowcolour

\small{DPR~\cite{DBLP:conf/sigir/ZhuWC20}} & Zhu et al. & 2020 & &  & \cmark  & \cmark &   &  & \cmark & \cmark &  &  & \cmark &  &\cmark & & & & \cmark & \cmark  \\ [1ex]

\small{FAN~\cite{DBLP:journals/corr/abs-2006-16742}} & Wu et al. & 2020 & &  & \cmark  & \cmark & &  & \cmark & &  & \cmark & \cmark &  & \cmark & & & \cmark & \cmark & \cmark\\ [1ex]

\bottomrule
\end{tabular}}}
\end{table}

\paragraph{Accuracy of Recommendations.}
\updated{Adversarial attacks against RS models have primarily focused on the degradation of recommendation predictive performance. Thus, in this section we review research works that perform adversarial attacks to undermine the accuracy of recommendation models and measure the impact of the adversarial attack (and defense) via an accuracy evaluation metric.}



Looking at Table~\ref{tbl:adl_attack} globally, we note that adversarial personalized ranking (APR)~\cite{DBLP:conf/sigir/0001HDC18} by He et. al. was the first work that formally addressed AML to improve the robustness of BPR-MF. After this pioneering work, in the following years, a growing number of works have considered application of AML for different recommendation tasks. Another interesting observation is the co-occurrence of the attack type FGSM and defense model adversarial training (AdReg). In fact, the adversarial training procedure based on FGSM is the first defense strategy proposed by~\citet{DBLP:journals/corr/GoodfellowSS14} to train DNNs resistant to adversarial examples. The authors interpret the improvement in robustness to adversarial examples because the proposed procedure is based on the minimization of the error on adversarially perturbed data.

\begin{table}[!t]
\caption{Evaluation and domain comparison of adversarial machine learning approaches for attack and defense on RS  (ML: Movielens, FL: FilmTrust, EM: EachMovie, CD: CiaoDVD, Yelp: YE, PI: Pinterest, AM: Amazon, 30M: 30Music , YA: Yahoo, AotM: Art of the Mix, F: Foursquare, T: Tradesy.com)}
\label{tbl:adl_evaluation}
{\scalebox{0.70}{\begin{tabular}{p{1.70cm} c c|p{0.13cm} p{0.13cm} p{0.13cm} p{0.13cm} p{0.13cm} p{0.13cm} p{0.13cm} p{0.13cm} p{0.13cm} | p{3.5cm} p{4cm}}
\toprule

\multicolumn{1}{c}{\textbf{Name}}
& \multicolumn{1}{c}{\textbf{Authors}}
& \multicolumn{1}{c}{\textbf{Year}} 
& \multicolumn{9}{c}{\textbf{Evaluation}} 
& \multicolumn{2}{c}{\textbf{Domain \& Dataset}} \\ 

\cmidrule(r){1-1}\cmidrule(r){2-2}\cmidrule(r){3-3}\cmidrule(r){4-12}\cmidrule(r){13-14}

&
&
& \multicolumn{2}{c}{\begin{tabular}[c]{@{}c@{}}pref. \\ type\end{tabular}} 
& \multicolumn{7}{c|}{\begin{tabular}[c]{@{}c@{}}evaluation \\ metric \end{tabular}}
&
\multicolumn{1}{l}{domain}
& \multicolumn{1}{l}{datasets} \\

\cmidrule(lr){4-5}\cmidrule(lr){6-12}
&
&
&\rotatebox{90}{\small{implicit}} &\rotatebox{90}{\small{explicit}}  &\rotatebox{90}{\small{NDCG}}  &\rotatebox{90}{\small{HR}} 
&\rotatebox{90}{\small{SR}} 
& \rotatebox{90}{\small{F1}} 
&\rotatebox{90}{\small{$L_2$-dist.}}
&\rotatebox{90}{\small{Precision}}
&\rotatebox{90}{\footnotesize{Pred. Shift}}
&  &
\\
\bottomrule
\multicolumn{3}{l|}{\textbf{Accuracy}} &  & & & & & & & &&   & \\ \hline 

\small{APR~\cite{DBLP:conf/sigir/0001HDC18}} &He at al. &2018 &\cmark & &\cmark &\cmark & & & & & &\small{tourism, SM/SN} & \small{YE, PI, GO}  \\ [1ex]
\myrowcolour

\small{ACAE~\cite{DBLP:conf/ijcnn/YuanYB19}} &Yuan et al. &2019 &\cmark & & &\cmark & & & & & &\small{movie}  &\small{ML 1M, CD, FT} \\ [1ex]

\small{FGACAE~\cite{DBLP:conf/sigir/YuanYB19}} &Yuan et al. &2019 &\cmark & & &\cmark & & & & & &\small{movie} &\small{ML 1M, CD, FT}\\ [1ex]
\myrowcolour

\small{AMR~\cite{8618394}} &Tang et al. &2019 &\cmark & &\cmark &\cmark & & & & & &\small{fashion}  &\small{PI, AM}   \\ [1ex]

\small{FNCF~\cite{DBLP:journals/tmm/DuFYXCT19}} &Du et al. &2019 &\cmark & &\cmark &\cmark &\cmark & &\cmark & & &\small{movie}  &\small{ML (100k, 1M)} \\
\myrowcolour

\small{ATF~\cite{DBLP:conf/recsys/ChenL19}} &Chen et al. &2019 & \cmark &  & & & &\cmark & & & &\small{movie, music}  &\small{ML, LastFM}  \\ [1ex]

\small{GANAtt~\cite{DBLP:conf/recsys/Christakopoulou19}} & \small{Christakopoulou} et al. & 2019 & &\cmark  & & & \cmark & & & & &\small{movie} & ML 1M \\ [5pt]
\myrowcolour

\small{AdvIR~\cite{DBLP:conf/www/ParkC19}} &Park et al. &2019 & \cmark & & & & & & &  \cmark & &\small{movie}& ML 100K  \\ [1ex]

\small{AMASR~\cite{DBLP:conf/sigir/TranSL19}} &Tran et al. &2019  & \cmark & &\cmark &\cmark & & & & & &\small{music}& 30M, AotM  \\ [1ex]
\myrowcolour

\small{ATMBPR~\cite{DBLP:journals/access/WangH20a}} &Wang et al. &2020 & \cmark & &\cmark &\cmark & & & &\cmark & & \small{tourism, SM/SN, movie}  & ML (100k, 1M), YA, YE, PI  \\ [1ex]

\small{SACRA~\cite{DBLP:conf/wsdm/LiW020}} &Li et al. &2020 & \cmark & & & & & & & \cmark & & \small{tourism, SM/SN, business}  &  YE, FS  \\ [1ex]
\myrowcolour

\small{MSAP~\cite{anelli2020multi}} & Anelli et al. & 2020  & \cmark & & \cmark & \cmark & & & & \cmark & &\small{movie, music} 
& ML-1M, LastFM \\ [0.5ex]

\small{TAaMR~\cite{DBLP:conf/dsn/DiNoia20}} & Di Noia et al. & 2020 & \cmark & & & \cmark &  & & & & &\small{fashion} 
& AM (Women, Men)   \\ [0.5ex] \myrowcolour

\small{AIP~\cite{DBLP:journals/corr/abs-2006-01888}} & Liu et al. & 2020 & \cmark & & & \cmark &  & & & & &\small{fashion} 
& AM Men, T \\ [0.5ex]

\small{SAO~\cite{DBLP:conf/sigir/ManotumruksaY20}}   & \small{Manotumruksa} et al. & 2020 & \cmark & & \cmark  & \cmark &  & & & & & \small{tourism, fashion, movie} & AM, ML-1M, F, Y  \\ [0.5ex]  \myrowcolour

\small{VAR~\cite{anelli2020empirical}} & Anelli et al. & 2020  & \cmark & & \cmark & \cmark &  & & & & &\small{fashion} 
& AM (Women, Men), T\\ [0.5ex]

\small{AUSH~\cite{DBLP:conf/cikm/LinC0XLY20}} & Lin et al. & 2020 & &\cmark  & & \cmark &  & & & & \cmark &\small{movie, automotive} & ML 100K, AM Auto, FT \\ [5pt]

\hline
\multicolumn{3}{l|}{\textbf{Privacy}} &  & & & & & & & &&   & \\ \myrowcolour \hline 

\small{PAT~\cite{DBLP:conf/icpram/ResheffESS19}} & Resheff et al. & 2019 & \cmark & & \cmark & &  & & & & &\small{movie} & ML 1M \\ [1ex]

\small{RAP~\cite{DBLP:conf/wsdm/BeigiMGAN020}} & Beigi et al. & 2020 & \cmark & & \cmark & &  & & & \cmark & &\small{movie} & ML 100k \\ [1ex]
\hline 

\multicolumn{3}{l|}{\textbf{Bias, Fairness}} &  & & & & & & & &&   & \\ \myrowcolour \hline 

\small{DPR~\cite{DBLP:conf/sigir/ZhuWC20}} & Zhu et al. & 2020 & \cmark & & \cmark & & \cmark  & & & & &\small{movie, business, product } & ML 1M, Y, AM \\ [1ex]

\small{FAN~\cite{DBLP:journals/corr/abs-2006-16742}} & Wu et al. & 2020 & \cmark & & \cmark & & & \cmark & & & &\small{news } & MSN-news \\ [1ex]

\bottomrule
\end{tabular}}}
\end{table}

Furthermore, in Table~\ref{tbl:adl_evaluation}, we provide an overview of the presented approaches under the perspective of experimental evaluation. In particular, we classify the surveyed works according to the \textit{preference score} used for building/training the recommender models according to implicit and explicit (i.e., rating-based) feedbacks, the prominent \textit{evaluation metrics} utilized for the offline evaluation of attack success (NDCG, HR, Success Rate, F1, distortion, Precision, and MAP), the \textit{domain} of focus (e.g., movie, music, social media, business) and \textit{datasets} used for evaluation. We may notice that, most of the approaches have been tested on an \textit{implicit} preference type. As for the evaluation metrics, HR is the most adopted one followed by nDCG with a partial overlap among approaches adopting them both. As for the application domain of the datasets used for the evaluation, \textit{movie} is the most adopted one. This is mainly due to the popularity the Movielens datasets (in their two variants 1M and 100k). Interestingly, \textit{tourism} is an emerging domain thanks to the availability of the Yelp dataset. Finally, we observe that the high majority of the baselines are based on MF approaches. The following section will provide a detailed description of the most prominent approaches.

\setlength{\parindent}{15pt} [\textbf{APR}] \citet{DBLP:conf/sigir/0001HDC18} are the first to propose an adversarial learning framework for recommendation. The proposed model, called \textit{adversarial personalized ranking (APR)}, examines the robustness of BPR-MF to adversarial perturbation on users and items embedding of a BPR-MF~\cite{DBLP:conf/uai/RendleFGS09}. The authors verify the success of using adversarial training as a defense strategy against adversarial perturbations and demonstrate the competitive results in applying adversarial training on BPR-MF. \updated{Recently,~\citet{anelli2020multi} studied application of iterative FGSM-based perturbation techniques and demonstrated the inefficacy of defacto defense mechanism APR in protecting the recommender against such multi-step attacks. For instance, the authors showed that the defended model loses more than 60\% of its accuracy under iterative perturbations, while only less than 9\% in the case of FGSM-ones.} 
\\\setlength{\parindent}{15pt} [\textbf{AMR}] \citet{8618394} put under adversarial framework another BPR model, namely visual-BPR (VBPR). VBPR is built upon BPR and extends it by incorporating visual dimensions (originally based on deep CNN feature) by using an embedding matrix. In~\cite{8618394}, the authors first motivate the importance for adversarial training of VBPR by visually depicting how a surprisingly modest amount of adversarial perturbation ($\epsilon = 0.007$) added on raw image pixels \textemdash~where the added noise is barely perceivable to the human eye \textemdash~can alter recommendation raking outcomes of VBPR and produce erroneous results. The proposed model therefore consists of constructing adversarial perturbations under the FGSM attack model and adding them to the deep latent feature of items' images extracted by CNN (i.e., ResNet50~\cite{DBLP:conf/cvpr/HeZRS16}) with the goal to learn robust image embedding parameters. One of the key insights about this work is that it does not add perturbations directly on raw image pixels for two main reasons: (i) it would require the feature extractor (CNN) component and the recommender model to be trained end-to-end with overfitting issues on the CNN due to the sparsity of user-item feedback data, (ii) it would be a time-consuming operation because at each update of the recommender model it is necessary to update all the CNN parameters.

In the above-mentioned works, the authors adopt several steps to validate the effectiveness of the proposed adversarial training framework, which can be summarized according to the following dimensions: (i) the \textit{generalization} capability, (ii) the comparison of \textit{adversarial noise v.s. random noise}, and (iii) the \textit{robustness of models}. Regarding (i), the key insight is that adversarial training approaches (i.e., APR and AMR) can lead to learning model parameters, which can enhance model generalization capability \textemdash~in other words, improvement of the general performance of recommendation while not being exposed to adversarial perturbation. Concerning (ii), it has been demonstrated that the impact adversarial perturbation on classical recommendation models (e.g., MF-BPR or VBPR) is significantly larger than their random noise counter-part under similar perturbation level. For instance,~\cite{8618394} shows that by exposing MF to adversarial and random noise, the test on nDCG is decreased by -21.2\% and -1.6\% respectively \textemdash~i.e., an impact of approximately 13 times difference. Dimension (iii) constitutes the core of the system validations in these works in which compelling evidence has been provided on the vulnerability of classical recommendation models to adversarial examples, or equivalently the robustness of the proposed training framework against adversarial samples. To provide an illustrating example, in~\cite{8618394} it has been shown for an experiment on the Amazon dataset, that by changing the perturbation level from $\epsilon = 0.05$ to $\epsilon = 0.2$, the amount of decrease in nDCG ranges from -8.7\% to -67.7\% whereas for AMR it varies from -1.4\% to -20.2\%. These results suggest that approaches using adversarial learning instead of classical learning act significantly in a more robust way against adversarial perturbations.\\
\setlength{\parindent}{15pt} [\textbf{AdvIR}] In~\cite{DBLP:conf/www/ParkC19}, the authors propose a system to address CF recommendation based on implicit feedbacks. The main issue in learning from implicit interaction is characterized by scarcity of negative feedbacks compared with positive ones, regarded as one-class problem. Sampling uniformly from unobserved data, known as \textit{negative sampling }, has been introduced in prior work to address this issue. The proposed system in~\cite{DBLP:conf/www/ParkC19} is called AdvIR, which entails an adversarial sampling and training framework to learn recommendation models from implicit interactions. The system applies adversarial training on both positive and negative interaction separately, to create informative adversarial positive/negative samples. The proposed adversarial training approach works for both discrete and continuous input by adding the adversarial perturbation directly on the input vector (e.g., one-hot encoding user-id).\\
\setlength{\parindent}{15pt} [\textbf{ACAE / FG-ACAE}] \citet{DBLP:conf/sigir/YuanYB19, DBLP:conf/ijcnn/YuanYB19} use the adversarial training framework for a neural network-based recommendation model, namely collaborative denoising auto-encoder (CDAE)~\cite{DBLP:conf/wsdm/WuDZE16}, based on which the authors propose two variations, namely: i) the adversarial collaborative auto-encoder (ACAE) and (ii) fine-grained collaborative auto-encoder (FG-ACAE). ACAE applies adversarial noise on encoder and decoder parameters and adopts an adversarial training framework. FG-ACAE considers the impact of adversarial noise in a more fine-grained manner. In particular, in FG-ACAE adversarial noise is added not only on encoder and decoder but also on the user's embedding matrix as well as hidden layers of the network. Furthermore, to increase the flexibility of training, all the noise factors in ACAE and FG-ACAE are controlled by different parameters. The experimental results confirm the trend that AdReg may improve the model's robustness against adversarial perturbed input, as well as the generalization performance of recommenders.\\
\setlength{\parindent}{15pt} [\textbf{ATF}] \citet{DBLP:conf/recsys/ChenL19} combine tensor factorization and adversarial learning to improve the robustness of pairwise interaction tensor factorization (PITF)~\cite{DBLP:conf/wsdm/RendleS10} for context-aware recommendation. Comparison with standard tensor models in tag recommendations acknowledges that the adversarial framework outperforms state-of-the-art tensor-based recommenders.\\
\setlength{\parindent}{15pt} [\textbf{FNCF}] \citet{DBLP:journals/tmm/DuFYXCT19} approach security issues for C\&W  attacks~\cite{DBLP:conf/sp/Carlini017}. The authors propose to make more robust neural network-based collaborative filtering models (e.g., NCF~\cite{DBLP:conf/www/HeLZNHC17}) by using knowledge distillation~\cite{DBLP:journals/corr/HintonVD15} instead of the adversarial (re)training. The framework integrates  knowledge distillation with the injection of additive adversarial noise at training time. Experiments demonstrate that this system enhances the robustness of the treated recommender model.\\
\setlength{\parindent}{15pt} [\textbf{SACRA}] \citet{DBLP:conf/www/LiLWZW19} propose a novel recommender model, named Click Feedback-Aware Network (CFAN), to provide query suggestions considering the sequential search queries issued by the user and her history of clicks.
The authors employ additional adversarial (re)training epochs (i.e., adding adversarial perturbations on item embeddings) to improve the robustness of the model. \updated{A similar approach has been also implemented by~\citet{DBLP:conf/sigir/ManotumruksaY20} to robustify a self-attention sequential recommender model, named [\textbf{SAO}].}\\
\setlength{\parindent}{15pt} [\textbf{TAaMR}] \citet{DBLP:conf/dsn/DiNoia20} explore the influence of targeted adversarial attacks (i.e., FGSM\cite{DBLP:journals/corr/GoodfellowSS14}, and PGD~\cite{DBLP:conf/iclr/MadryMSTV18}) against original product images used to extract deep features in state-of-the-art visual recommender models (i.e., VBPR~\cite{DBLP:conf/aaai/HeM16}, and AMR~\cite{8618394}). The authors verify that recommendation lists can be altered such that a low recommended product category can be pushed by adding adversarial noise on product images in a human-imperceptible way.   \updated{Within a similar scenario,~\citet{anelli2020empirical} verify, in the \textbf{VAR} framework, the inefficacy of the adversarial robustification of the image feature extractor component in protecting the visual recommender from such adversarial perturbations}.\\
\updated{
\setlength{\parindent}{15pt} [\textbf{AIP}] Similar to TAaMR~\cite{DBLP:conf/dsn/DiNoia20},~\citet{DBLP:journals/corr/abs-2006-01888} propose a series of adversarial attacks to increase the recommendability of items by perturbing the products images. The authors model three level of adversary's knowledge (i.e., high, medium, and low) with corresponding adversarial image perturbation strategies. Furthermore, they validate the inefficacy of JPEG compression and bit depth reduction as two possible defense mechanisms.}



\paragraph{Users' Privacy.}
\updated{Another main area of concern is user privacy and averting the negative consequences of adversarial attacks on user privacy. Recently, in the light of privacy-violation scandals such as Cambridge Analytica~\cite{DBLP:conf/www/GonzalezYFLA19} privacy-protection laws such as the GDPR, US Congress, and other jurisdictions have been proposed to legislate new disclosure laws. Thus, attempts have been made to build machine-learned recommendation models that offer a privacy-by-design architecture, such as federated learning~\cite{DBLP:conf/aiia/AnelliDNF19}, or the ones based on differential privacy~\cite{DBLP:conf/icalp/Dwork06}. Nonetheless, several works recently have challenged user data confidentiality via adversarial attacks, for example, in the context of \textit{social recommenders} that has been widely studied in these scenarios. For instance,~\citet{DBLP:conf/aaai/MengWSLCLZ18, DBLP:journals/www/MengWSLCLZ19} propose a privacy-preserving framework to protect users from adversaries that want to infer, or reconstruct, their historical interactions and social connections.~\citet{DBLP:journals/tkde/YangQC19} propose to defend users' privacy from inference attacks with a privacy-oriented social media data publishing framework optimized to preserve the recommendation performance, while domain-independent recommendation algorithms have been developed by~\citet{DBLP:journals/tkde/ShinKSX18} as a MF extension. All these works use the differential privacy~\cite{DBLP:conf/icalp/Dwork06} technique to reduce privacy violations.}


\updated{
\setlength{\parindent}{15pt} [\textbf{PAT}]~\citet{DBLP:conf/icpram/ResheffESS19} propose an adversarial training procedure, the Domain-Adversarial Training~\cite{DBLP:series/acvpr/GaninUAGLLML17}, to build a privacy-adversarial method to defeat the data leakage. The intuition is to train a LFM with the classical minimax paradigm where the model learns its parameters by minimizing both the recommendation cost an adversarial regularization component related to the adversarial privacy-violation.   
}\\
\updated{
\setlength{\parindent}{15pt} [\textbf{RAP}]~\citet{DBLP:conf/wsdm/BeigiMGAN020} propose an adversarial learning procedure to protect users' from attribute-inference attacks. The model, named Recommendation with Attribute Protection (RAP), simultaneously learns to maximize the users' gain from the recommendation while minimizing the adversary capability in inferring users' personal attributes (e.g., gender, age, and occupancy).}

\paragraph{Bias, Fairness.} \updated{Another area of concern is related to biases and fairness of recommendations. RS assist users in many life-affecting scenarios such as medical, financial, or job-related ones. Unfair recommendations could have far-reaching consequences, impacting people's lives and placing minority groups at a significant disadvantage~\cite{verma2018fairness,speicher2018unified}. From a RS perspective, where users are first-class citizens, fairness is a \textit{multi-sided concept} and the utility of recommendations needs to be studied by considering the benefits of multiple groups of individual~\cite{deldjoo2020flexible}, for instance based on the user-centered utility and the vendor-centered utility (e.g., profitability). In the literature, a few research works have exploited the adversarial training procedure to reduce the biased/unfair impact of recommendations.} 
\\\updated{\setlength{\parindent}{15pt} [\textbf{DPR}] 
Inspired by~\cite{DBLP:conf/nips/LouppeKC17},~\citet{DBLP:conf/sigir/ZhuWC20} propose a debiased personalized ranking (DPR) model composed of two components: an adversary model, i.e., multi-layer perceptron network, and a classical recommender model, i.e., BPR-MF.  During the adversarial training, the adversary tries to infer the group of an item, while the recommender model tries to reduce both the recommendation error and the adversary's capability of identifying the true class of the item. The central intuition is to unbias the recommender model by enhancing the similarity of the predicted score distributions between different item groups. Extensive experiments demonstrate that DPR reduces the under-recommendation bias while retaining accurate recommendations.} \\
\updated{
\setlength{\parindent}{15pt} [\textbf{FAN}] \citet{DBLP:journals/corr/abs-2006-16742} design a fairness-aware news recommendation (FAN) method via adversarial learning. Similar to DPR~\cite{DBLP:conf/nips/LouppeKC17}, the authors extend the neural news recommendation with a multi-head self-attention (NRMS) model~\cite{DBLP:conf/emnlp/WuWGQHX19} with an adversarial component trained to infer the sensitive users' characteristics, while the recommender model is regularized to reduce the adversarial possibility of inferring the users' attributes. The intuition is that adversarial training generates bias-free users' embeddings that can be in turn used to produce fair-aware recommendations.}

\section{Adversarial Learning for GAN-based Recommendation}~\label{sec:GAN}
What we presented in Section~\ref{sec:security} deals with the class of \dquotes{discriminative} models where the main aim is to learn the conditional probability $p(y|x)$. 
The focus of the current section is on a novel class of \dquotes{generative} models, named \textit{Generative Adversarial Networks} (GANs). 
Loosely speaking, a generative model cares about the \textit{generative} process behind data ---or product features in a recommendation scenario ---  to categorize the data instances. Here the focus is on learning $p(x|y)$ from the data.


GANs are a powerful class of generative models that use two networks ---trained simultaneously in a zero-sum game--- with one network focused on data generation and the other one centered on discrimination. The adversarial learning scheme --- or the min-max game --- which lies in the heart of GANs empowers these ML models with phenomenal capabilities such as the ability to model high-dimensional distributions. As a result, these networks have been exploited to solve challenging problems in computer vision. 
The research in RS community has used the generalization (or in technical term data distribution capturing) potential of GANs as an opportunity to solve a variety of tasks relevant to RS.

As it can be noted the term \dquotes{adversarial} inside generative adversarial networks refers to the learning scheme used by these models and \textit{not the application}. In other words, the application of GANs for RS covers variety of aspects not limited to the security of RS, as we will see in the subsequent sections.

\updated{In the following, we present the foundations of GANs in Section~\ref{subsec:found_GAN}. Section~\ref{subsec:GAN_RS} provides a conceptual framework on applications of GANs in RS. Finally, in Section~\ref{subsec:GAN_appl_RS} we focus on the new trends of GAN research in RS and provide a detailed literature review on this topic by categorizing the research works according to collaborative (cf. Section~\ref{subsec:GAN_CF_rec}), context-aware (cf. Section~\ref{subsec:GAN_CA_rec}), and cross domain (cf. Section~\ref{subsec:GAN_CR_rec})}.

\begin{figure}[t]
    \includegraphics[width=0.85\textwidth]{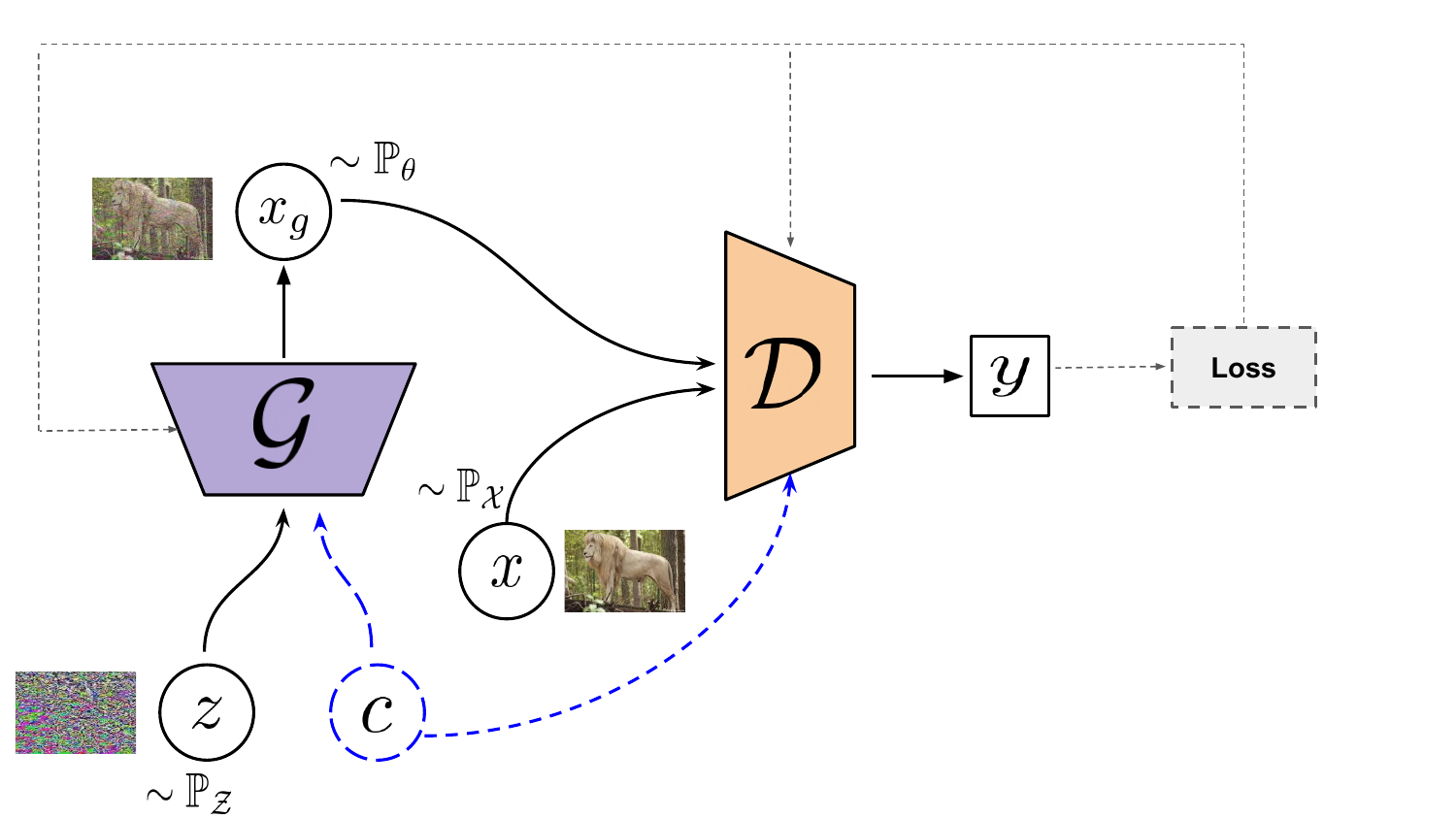}
  \caption{Schematic comparison of two well-known GAN models: (a) conventional vanilla GAN with filled color and (b) Conditional GAN (which includes the dashed blue entities)}
  \label{fig:gan_architecture}
\end{figure}
\subsection{Foundations of Generative Adversarial Networks (GANs)}\label{subsec:found_GAN}

GANs are deep generative models proposed by~\citet{DBLP:journals/corr/GoodfellowPMXWOCB14} in 2014. A GAN is composed of two components, a generator $\mathcal{G}$, and a discriminator $\mathcal{D}$. The generator works to capture the real data distribution to generate adversarial examples and fool the discriminator,  while the discriminator endeavors to distinguish the fake examples from real ones.
This competition, known as adversarial learning, ends when the components reach the Nash equilibrium. The GAN architecture is shown in Figure~\ref{fig:gan_architecture}.

\begin{definition}[Conventional Vanilla GAN~\citet{DBLP:journals/corr/GoodfellowPMXWOCB14}]\label{def:vanilla-GAN}
Assume that we are given a dataset of input samples $x \in \mathcal{X}$, where $\mathbb{P}_{\mathcal{X}}$ represents the probability distribution of the original data and suppose $z \in \mathcal{Z}$ denotes a sample from some latent space $\mathcal{Z}$. We are interested in sampling from $\mathbb{P}_{\mathcal{X}}$. The goal of GAN is to train the generator $\mathcal{G}$ to transform samples $z \sim \mathbb{P}_{\mathcal{Z}}$ into $g_{\theta}(z) \sim \mathbb{P}_{\theta}$ such that $ \mathbb{P}_{\theta} \approx \mathbb{P}_{\mathcal{X}}$. The role of the discriminator $\mathcal{D}$ is to distinguish $\mathbb{P}_{\theta}$ and $\mathbb{P}_{\mathcal{X}}$ by training a classifier $f_{\phi}$. The training involves solving the following min-max objective
    \begin{equation}
         \label{eq:gan_basic}
         \min_{\theta} \max_{\phi} L(\mathcal{G}_\theta, \mathcal{D}_{\phi}) =  \, \mathbb{E}_{x \sim \mathbb{P}_{\mathcal{X}}} \, \log f_{\phi}(x) + \mathbb{E}_{z \sim \mathbb{P}_{\mathcal{Z}}} \log (1-f_{\phi}(g_{\theta}(z)))
    \end{equation}
where $\theta$ and $\phi$ are model parameters of the discriminator and generator respectively, learned during the trained phase.  \qed 
\end{definition}
Different distance measures $f_{\theta}$ lead to different GAN models, e.g., Vanilla GAN (based on Jensen-Shannon divergence)~\cite{DBLP:journals/corr/GoodfellowPMXWOCB14}, Wasserstein GAN (based on Wasserstein distance)~\cite{DBLP:journals/corr/ArjovskyCB17}, and Conditional GAN (based on class conditioning on both the generator and discriminator)~\cite{DBLP:journals/corr/MirzaO14}.

\begin{definition}[Conditional-GAN (CGAN)\cite{DBLP:journals/corr/MirzaO14}]
Conditional GAN extends the conventional GAN by incorporating an extra condition information term $c$ on both the input of the generator $\mathcal{G}$ and the discriminator $\mathcal{D}$, thus conditioning them on this new term
    \begin{equation}
         \label{eq:gan_basic}
         \min_{\theta} \max_{\phi} L(\mathcal{G}_\theta, \mathcal{D}_{\phi}) =  \, \mathbb{E}_{x \sim \mathbb{P}_{\mathcal{X}}} \, \log f_{\phi}(x|c) + \mathbb{E}_{z \sim \mathbb{P}_{\mathcal{Z}}} \log (1-f_{\phi}(g_{\theta}(z|c)))
    \end{equation}
where $c$ can represent any auxiliary information to the networks such as class labels, content features, data from other domains and so forth. \qed
\end{definition}

\subsection{GAN-based Recommendation Framework}\label{subsec:GAN_RF}

GANs have been successfully applied in start-of-the-art RS to learning recommendation models. Since the first pioneering GAN-based work IRGAN~\cite{DBLP:conf/sigir/WangYZGXWZZ17} in 2017, we have witnessed rapid adoption of these network architectures in many traditional and novel applications and domains. In this section, we provide a conceptual framework that will show how GANs are employed in RS domain and shed light on particularities and differences of GAN application in RecSys and ML. 

\begin{figure}[!h]
    \centering
    \includegraphics[width=0.85\textwidth]{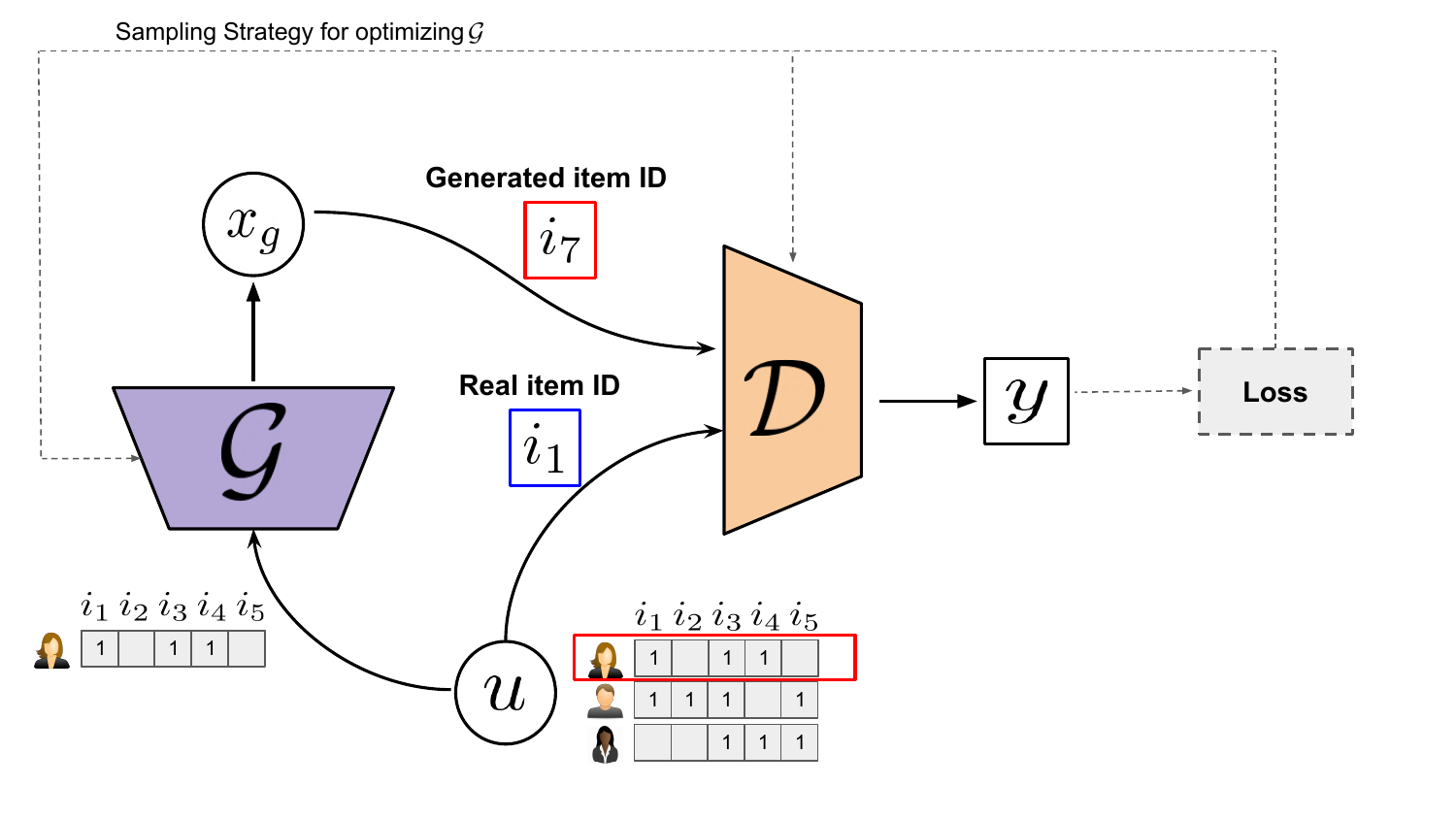}
    \caption{A conceptual view of GAN-CF incorporating GAN to address item recommendation task.}
    \label{fig:GAN-CF}
\end{figure}

\noindent {\textbf{GAN-CF problem formulation and conceptual model.}}\label{subsec:GAN_RS}
The prominent recommendation models in the literature that successfully apply GAN~\cite{DBLP:conf/sigir/WangYZGXWZZ17,DBLP:conf/aaai/WangWWZZZXG18} for the CF task, utilize the two-player min-max game with objective function built on top of Eq.~\ref{eq:gan_basic}.
\begin{definition}[The GAN-CF model~\cite{DBLP:conf/sigir/WangYZGXWZZ17}]
Let $\mathcal{U}$ and $\mathcal{I}$ denote a set of users and items in a system, respectively. The training objective is given by
    \begin{equation}
         \label{eq:irgan_basic}
          \min_{\theta} \max_{\phi} L(\mathcal{G}_\theta, \mathcal{D}_{\phi}) = \, \mathbb{E}_{i \sim \mathbb{P}_{\mathcal{X}}(i|u)} \, \log f_{\phi}(i|u) + \mathbb{E}_{\hat{i} \sim \mathbb{P}_{\theta}(\hat{i}|u)} \log \, (1-f_{\phi}(\hat{i} | u))
    \end{equation}
where $i \in \mathcal{I}$ is an item receiving implicit (or explicit) feedback by user $u \in \mathcal{U}$ (e.g., purchased) and $\hat{i} \in \mathcal{I}$ is a generated item.
\qed 
\end{definition}
A few observations are important to be made here: (i) the output of generator $\mathcal{G}$ is a set of item indices deemed relevant to user $u$; (ii) both $\mathcal{G}$ and $\mathcal{D}$ are \textit{user-conditioned}, signifying that model parameters are learnt in a \textit{personalized fashion}; (iii) the GAN-based CF works do not use the noise term as input (to $\mathcal{G}$) as the goal is to generate one unique ---yet plausible--- item rather than a set of items. Figure~\ref{fig:GAN-CF} summarizes these aspects conceptually.

\begin{table}[!h]
\caption{Key sampling strategies proposed for CF-GAN recommendation models}

\resizebox{0.9\textwidth}{!}{%

\label{tbl:smple_strg}
\centering
\begin{tabular}{@{} lll @{}} 
\toprule
    \textbf{Method} & 
    \textbf{Key insight} & 
    \textbf{Formal description} \\  
    \midrule
    \begin{tabular}[l]{@{}l@{}}
    \small{REINFORCE*} \\ 
    \small{\cite{DBLP:conf/nips/SuttonMSM99}}
    \end{tabular} & 
    \small{
    \begin{tabular}[l]{@{}l@{}}Optimize $\mathcal{G}$ with $K$ discrete \\ items $\hat{i}_k$.\end{tabular}}
     & 
    \begin{minipage}{7cm}
    $ \nabla_{\theta} \simeq \frac{1}{K} \sum_{k=1}^{K} \nabla_{\theta} \log \mathbb{P}_{\theta}(\hat{i}_k | u,r) \log (1-f_{\phi}(\hat{i}_k | u)))$
    \end{minipage} \\ [5pt] 
    \begin{tabular}[l]{@{}l@{}}
    \small{Gumbel-Softmax**} \\\small{\cite{DBLP:conf/iclr/JangGP17, DBLP:journals/eswa/SunWWY19}}
    \end{tabular}
    & 
    \small{
    \begin{tabular}[l]{@{}l@{}}Approximate discrete items \\ with virtual items $v_k$ through \\ a differentiable estimator. \end{tabular}}
    & \begin{minipage}{7cm}
    $ v_k = \frac{\exp \left(\left(\log \mathbb{P}_{\theta}\left(i_{k}| u, r\right)+g_{k}\right) / \tau\right)}{\sum_{j=1}^{K} \exp \left(\left(\log \mathbb{P}_{\theta}\left(i_{j}| u, r\right)+g_{j}\right) / \tau\right)}$
    \end{minipage} \\ 
    \midrule
    \multicolumn{3}{l}{
    \small{
    \begin{tabular}[l]{@{}l@{}}
     \textit{* $\nabla_{\theta}$ is the gradient of the generator $\mathcal{G}$.} \\
     \textit{**In the Gumbel-Softmax formulation $g_k$ and $g_j$ represent sampled noise, and $\tau$ is a temperature hyper-parameter} \\ 
        \textit{to control the smooth of distribution ($\tau \simeq 0$, the probability is concentrated to few items).}
    \end{tabular}
    }
    } \\ \bottomrule
\end{tabular}

}
\end{table}
\noindent {\textbf{Discrete outcome and sampling strategies.}} 
The parameters in the GAN-CF model are learned in an end-to-end fashion. 
However, before we can take benefit of this training paradigm, the system needs to solve a critical issue that does not exist on the original GAN presented in Def. 3.1. based on the sampled noise signal. The generation of recommendation lists is a discrete sampling operation, i.e., performed over discrete candidate items (see Figure~\ref{fig:GAN-CF}). Thus, the gradients that are derived from the objective function in Eq.~\eqref{eq:irgan_basic} cannot be directly used to optimize the generator via gradient descent as happens in the original GAN formulation, where gradients are applied for differentiable values (e.g.,images and videos). To obtain a differentiable sampling strategy in GAN-CF models, two sampling strategies are proposed in the literature based on reinforcement learning algorithm and the Gumbel-Softmax differentiable sampling procedure~\cite{DBLP:conf/nips/SuttonMSM99,DBLP:conf/iclr/JangGP17, DBLP:journals/eswa/SunWWY19}, summarized in Table~\ref{tbl:smple_strg}.

\subsection{GAN-based Recommendation Models: State of the Art}\label{subsec:GAN_appl_RS}
We have identified a total of \updated{53 papers}  that integrate GAN in order to accomplish a particular RS-related task, and we classified them according to:

\begin{enumerate}
    \item Collaborative Recommendation
    \item Context-aware Recommendation
    \item Cross-domain Recommendation
    \item Fashion Recommendation
\end{enumerate}

We present Table~\ref{tbl:gan-approaches} to summarize the proposed models and provide insights about the constituting building blocks of the GAN model. From a global perspective, we can see a correlation between the class of $\mathcal{G}$, $\mathcal{D}$ and the recommendation task. For example, recursive models based on RNN are used for CA \textbf{\textit{Temporal-aware Rec.}} tasks, areas where these models can better capture the sequence information. This is while, for \textbf{\textit{Collaborative Rec.}} tasks, the rest of models are commonly used (e.g., Linear LFM, MLP and so on). It is interesting to note that CNN is used for majority of works in \textbf{\textit{Fashion Rec.}} From a training perspective, we can see that both point-wise and pair-wise models are almost equally used in all these works, perhaps indicating the point-wise training is still a useful method for evaluation of many GAN-based related RS models. In the following, we review each of these application scenarios by describing the most prominent approaches.
\vspace{-2mm}

\subsubsection{Collaborative Recommendation}\label{subsec:GAN_CF_rec}
GANs have been shown powerful in generating relevant recommendations --- in particular, using the CF approach --- and capable of successively competing with state-of-the-art models in the field of RS. We have identified the following reasons for the potential of GANs in RS: (i) they are able to generalize well and learn unknown user preference distributions and thus be able to model user preference in complex settings (e.g., IRGAN~\cite{DBLP:conf/sigir/WangYZGXWZZ17} and CFGAN~\cite{DBLP:conf/cikm/ChaeKKL18}); (ii) they are capable of generating more negative samples than random samples in pairwise learning tasks (e.g., APL~\cite{DBLP:journals/eswa/SunWWY19}, DASO~\cite{DBLP:conf/ijcai/FanD0WTL19}) and (iii) they can be used for data augmentation (e.g., AugCF~\cite{DBLP:conf/kdd/WangYWNHC19}  and RAGAN~\cite{DBLP:conf/www/ChaeKKC19}).\\
\setlength{\parindent}{15pt} [\textbf{IRGAN}] The work by Wang et. al.~\cite{DBLP:conf/sigir/WangYZGXWZZ17} is presumably the first attempt to integrate the generative and discriminative approach to IR under the same roof by proposing a GAN-based IR model. The authors demonstrate the application of IRGAN for web search, item recommendation and question answering tasks where for the item recommendation task, the query is constructed from the user's historical interactions.
During adversarial learning ---the min-max game--- the generator learns the actual distribution of relevant items as much as possible.  It turns out that this novel training idea results in a more satisfactory accuracy in recommendation than optimizing the traditional pure discriminative loss functions based on pointwise, or pairwise, objectives.
\begin{table}[]
\footnotesize
\centering
\caption{A schematic representation of GAN-based approaches to recommendation.}\label{tbl:gan-approaches}
\resizebox{0.83\textwidth}{!}{%
\begin{tabular}{ l c|lllllll|llllll|ll}
\toprule
\multicolumn{1}{c}{\textbf{Name}} & \textbf{Year} & \multicolumn{7}{c}{\textbf{Generator } ($\mathcal{G}$)} & \multicolumn{6}{c}{\textbf{Discriminator ($\mathcal{D}$)}} & \multicolumn{2}{c}{\textbf{Training}} \\ 
\cmidrule(lr){1-1}\cmidrule(lr){2-2}\cmidrule(lr){3-9}\cmidrule(lr){10-15}\cmidrule(lr){16-17}
 &  & \rotatebox{90}{LFM} & \rotatebox{90}{MLP} & \rotatebox{90}{CNN} & \rotatebox{90}{AE} & \rotatebox{90}{VAE} & \rotatebox{90}{LSTM} & \rotatebox{90}{GRU} & \rotatebox{90}{LFM} & \rotatebox{90}{MLP} & \rotatebox{90}{CNN} & \rotatebox{90}{AE} & \rotatebox{90}{LSTM} & \rotatebox{90}{GRU} & \rotatebox{90}{point} & \rotatebox{90}{pair}\\ 
 
\hline
\multicolumn{2}{|c|}{\textit{\textbf{Collaborative Rec.}}} & \multicolumn{15}{l|}{} \\
\hline

IRGAN~\cite{DBLP:conf/sigir/WangYZGXWZZ17} & 2017 & \checkmark &  &  &  &  &  &  & \checkmark &  &  &  &  &  & \checkmark & \checkmark   \\ \myrowcolour
CFGAN~\cite{DBLP:conf/cikm/ChaeKKL18} & 2018 &  & \checkmark &  &  &  &  &  &  & \checkmark &  &  &  &  & \checkmark &    \\
Chae et al.~\cite{8525831} & 2018 &  &  &  & \checkmark &  &  &  & \checkmark &  &  &  &  &  &  & \checkmark   \\ \myrowcolour
AVAE~\cite{8663730} & 2018 &  &  &  &  & \checkmark &  &  & &  \checkmark  &  &  &  &  & \checkmark &    \\
GAN-VAE-CF~\cite{DBLP:conf/cikm/KrishnanSSS18} & 2018 &  &  &  & \checkmark & \checkmark &  &  & &  \checkmark  &  &  &  &  &  &  \checkmark  \\ \myrowcolour
CAAE~\cite{DBLP:journals/access/ChaeSK19} & 2019 &  &  &  & \checkmark &  &  &  & \checkmark &  &  &  &  &  &  & \checkmark   \\ 
CGAN~\cite{DBLP:conf/icde/TongLZSC19} & 2019 &  &  &  &  & \checkmark &  &  &  & \checkmark &  &  &  &  & \checkmark &    \\ \myrowcolour
CALF~\cite{DBLP:conf/flairs/CostaD19} & 2019 &  &  & \checkmark &  &  &  &  &  &  & \checkmark &  &  &  &  & \checkmark   \\
PD-GAN~\cite{DBLP:conf/ijcai/WuLMZZG19} & 2019 & \checkmark &  &  &  &  &  &  & \checkmark &  &  &  &  &  &  & \checkmark   \\ \myrowcolour
LambdaGAN~\cite{DBLP:conf/ijcnn/WangZC019} & 2019 & \checkmark &  &  &  &  &  &  & \checkmark &  &  &  &  &  &  & \checkmark   \\ 
VAEGAN~\cite{DBLP:conf/ijcai/YuZCX19} & 2019 &  &  &  &  & \checkmark &  &  &  & \checkmark &  &  &  &  & \checkmark &    \\ 
\myrowcolour
APL~\cite{DBLP:journals/eswa/SunWWY19} & 2019 & \checkmark &  &  &  &  &  &  & \checkmark &  &  &  &  &  &  & \checkmark   \\
RsyGAN~\cite{DBLP:conf/ijcnn/YinL0019} & 2019 &  &  &  & \checkmark &  &  &  &  & \checkmark &  &  &  &  & \checkmark &   \\ 
\myrowcolour
GAN-PW/LSTM~\cite{DBLP:conf/ijcnn/ChenZWW019} & 2019 &  &  &  &  &  & \checkmark &  &  & \checkmark &  &  &  &  & \checkmark &    \\ 
CoFiGAN~\cite{DBLP:journals/kbs/LiuPM20} & 2020& \checkmark &  &  &  &  &  &  & \checkmark &  &  &  &  &  & \checkmark & \checkmark   \\ 
\myrowcolour
GCF~\cite{DBLP:conf/cikm/YuanYB20} & 2020& \checkmark &  &  \checkmark &  &  &  &  & \checkmark & \checkmark &  &  &  &  & \checkmark & \checkmark   \\ 

\hline
\multicolumn{2}{l|}{\textbf{Graph-based Collaborative Rec.}} &  &  &  &  &  &  &  &  &  &  &  &  &  &  &   \\ 
\hline

\myrowcolour
GraphGAN~\cite{DBLP:conf/aaai/WangWWZZZXG18} & 2018 & \checkmark &  &  &  &  &  &  & \checkmark &  &  &  &  &  & \checkmark &    \\ 
GAN-HBNR~\cite{DBLP:conf/aaai/CaiHY18} & 2018 &  &  &  & \checkmark &  &  &  &  &  &  & \checkmark &  &  & \checkmark &    \\ \myrowcolour
VCGAN~\cite{DBLP:conf/ismis/ZhangYCD18} & 2018 &  &  &  &  & \checkmark  &  &  & \checkmark &  &  &  &  &  & \checkmark &    \\
UPGAN~\cite{DBLP:journals/corr/abs-2003-12718} & 2020 & & \checkmark  &  &  &  &  &  &  & \checkmark &  &  &  &  & \checkmark &   \\

\hline
\multicolumn{2}{l|}{\textbf{Hybrid Collaborative Rec.}} &  &  &  &  &  &  &  &  &  &  &  &  &  &  &    \\ \myrowcolour
\hline

VAE-AR~\cite{DBLP:conf/cikm/LeeSM17} & 2017 &  &  &  &  & \checkmark &  &  &  & \checkmark &  &  &  &  & \checkmark &    \\ 
RGD-TR~\cite{DBLP:conf/icdm/LiH018} & 2018 &  & \checkmark &  &  &  &  &  &  & \checkmark &  &  &  &  & \checkmark &    \\ \myrowcolour
aae-RS~\cite{DBLP:conf/webi/YiHQ18} & 2018 &  &  &  & \checkmark &  &  &  &  & \checkmark &  &  &  &  & \checkmark &    \\
SDNet~\cite{DBLP:journals/tois/ChenZXQZ19} & 2019 & \checkmark &  &  &  &  &  &  & \checkmark &  &  &  &  &  & \checkmark &   \\ \myrowcolour
ATR~\cite{DBLP:conf/sigir/RafailidisC19} & 2019 &  &  &  &  &  &  & \checkmark &  &  & \checkmark &  &  &  & \checkmark &    \\ 
AugCF~\cite{DBLP:conf/kdd/WangYWNHC19} & 2019 &  & \checkmark &  &  &  &  &  & \checkmark &  &  &  &  &  & \checkmark &   \\ \myrowcolour
RSGAN~\cite{DBLP:journals/corr/abs-1909-03529} & 2019 &  &  &  & \checkmark &  &  &  & \checkmark &  &  &  &  &  &  & \checkmark   \\
RRGAN~\cite{DBLP:conf/ijcnn/ChenZWW019} & 2019 & \checkmark &  &  &  &  &  &  &  & \checkmark &  &  &  &  & \checkmark &    \\ \myrowcolour
UGAN~\cite{DBLP:conf/pakdd/WangGWYWX19} & 2019 & \checkmark &  &  &  &  &  &  & \checkmark &  &  &  &  &  & \checkmark &    \\
LARA~\cite{DBLP:conf/wsdm/SunLLRGN20} & 2020 &  & \checkmark &  &  &  &  &   &  &  \checkmark  & &  &  &  & \checkmark &    \\   \myrowcolour
CGAN~\cite{DBLP:journals/access/ChonwiharnphanT20} & 2020 &  &  &  & \checkmark  &  &  &   &  &  \checkmark  & &  &  &  & \checkmark &   \\ 

\hline
\multicolumn{2}{|c|}{\textit{\textbf{Context-aware Rec.}}} & \multicolumn{15}{l|}{} \\ 
\hline

\multicolumn{2}{l|}{\textbf{Temporal-aware}} &  &  &  &  &  &  &  &  &  &  &  &  &  &  &   \\   \myrowcolour
\hline

RecGAN~\cite{DBLP:conf/recsys/BharadhwajPL18} & 2018 &  &  &  &  &  &  & \checkmark &  &  &  &  &  & \checkmark & \checkmark &    \\ 
NMRN-GAN~\cite{DBLP:conf/kdd/WangYHLWH18} & 2018 &  & \checkmark &  &  &  &  &  &  &  &  &  & \checkmark &  &  & \checkmark   \\ \myrowcolour
AAE~\cite{DBLP:conf/recsys/VaglianoGMS18} & 2018 &  &  &  & \checkmark &  &  &  &  & \checkmark &  &  &  &  & \checkmark &    \\ 
PLASTIC~\cite{DBLP:conf/ijcai/ZhaoWYGYC18} & 2018 & \checkmark &  &  &  &  & \checkmark &  & \checkmark &  &  &  & \checkmark &  &  & \checkmark   \\ \myrowcolour
LSIC~\cite{8643032} & 2019 & \checkmark &  &  &  &  & \checkmark &  & \checkmark &  &  &  & \checkmark &  &  & \checkmark   \\ 
GAN-CDQN~\cite{DBLP:conf/icml/Chen0LJQS19} & 2019 &  &  &  &  &  & \checkmark &  &  & \checkmark &  &  &  &  & \checkmark &    \\ \myrowcolour

AOS4Rec~\cite{DBLP:conf/ijcai/ZhaoSZXB20} & 2020 &  & \checkmark &  &  &  &  &  \checkmark &  & \checkmark &  &  &  &  \checkmark  & &  \checkmark    \\
MFGAN~\cite{DBLP:conf/sigir/RenLLZWDW20} & 2020 &  & \checkmark &  &  &  &  &  &  & \checkmark &  &  &  &  & &  \checkmark    \\

\hline
\multicolumn{2}{l|}{\textbf{Geographical-aware}} &  &  &  &  &  &  &  &  &  &  &  &  &  &  &    \\ \cline{1-2} \myrowcolour
\hline

Geo-ALM~\cite{DBLP:conf/ijcai/0061W0Y19} & 2019 & \checkmark &  &  &  &  &  &  & \checkmark &  &  &  &  &  &  & \checkmark   \\ 
APOIR~\cite{DBLP:conf/www/0002YZTZW19} & 2019 &  &  &  &  &  &  & \checkmark &  & \checkmark &  &  &  &  & \checkmark &    \\

\hline
\multicolumn{2}{|c|}{\textit{\textbf{Cross-domain Rec.}}} & \multicolumn{15}{l|}{} \\   \myrowcolour
\hline

VAE-GAN-CC~\cite{DBLP:journals/corr/abs-1812-06229} & 2018 &  &  &  &  & \checkmark &  &  &  & \checkmark &  &  &  &  & \checkmark &    \\
RecSys-DAN~\cite{8698453} & 2019 &  &  & \checkmark &  &  &  &  &  & \checkmark &  &  &  &  & \checkmark &    \\\myrowcolour
FR-DiscoGAN~\cite{DBLP:conf/bigcomp/JoJCJ19}  & 2019 &  &  &  \checkmark &  &  &  &  &  &  & \checkmark  &  &  &  &  & \checkmark   \\
DASO~\cite{DBLP:conf/ijcai/FanD0WTL19} & 2019 &  & \checkmark &  &  &  &  &  &  & \checkmark &  &  &  &  & \checkmark &    \\ \myrowcolour
CnGAN~\cite{DBLP:conf/www/PereraZ19} & 2019 &  &  &  & \checkmark &  &  &  &  & \checkmark &  &  &  &  &  & \checkmark   \\
Asr~\cite{DBLP:conf/cikm/KrishnanCTS19}  & 2019 &  &  \checkmark &  & \checkmark & \checkmark &  &  & \checkmark  & \checkmark & \checkmark &  &  &  &  & \checkmark   \\ 
\myrowcolour
ALTRec~\cite{DBLP:journals/jcst/LiXZFCZ20}  & 2020 &  & &  & \checkmark & &  &  & & \checkmark &  &  &  &  &  & \checkmark   \\ 

\hline 
\multicolumn{2}{|c|}{\textit{\textbf{Fashion Rec.}}} & \multicolumn{15}{l|}{} \\   \myrowcolour
\hline

DVBPR~\cite{DBLP:conf/icdm/KangFWM17} & 2017 &  &  & \checkmark &  &  &  &  &  &  & \checkmark &  &  &  & \checkmark &    \\
CRAFT~\cite{DBLP:conf/eccv/HuynhCTA18} & 2018 &  & \checkmark &  &  &  &  &  &  & \checkmark &  &  &  &  & \checkmark &    \\ \myrowcolour
MrCGAN~\cite{DBLP:conf/aaai/ShihCLS18} & 2018 &  &  & \checkmark &  &  &  &  &  &  & \checkmark &  &  &  &  & \checkmark   \\ 
Yang \textit{et al.}~\cite{Yang2018FromRT} & 2018 &  &  & \checkmark &  &  &  &  &  &  & \checkmark &  &  &  &  & \checkmark   \\ \myrowcolour
$c^+$GAN~\cite{DBLP:journals/corr/abs-1906-05596} & 2019 &  &  & \checkmark &  &  &  &  &  &  & \checkmark &  &  &  &  & \checkmark  \\ 
 \hline
 
\end{tabular}}
\end{table}
\\ \setlength{\parindent}{15pt} [\textbf{GraphGAN}] \citet{DBLP:conf/aaai/WangWWZZZXG18} propose GraphGAN --- a graph-based representation learning --- (a.k.a. network embedding) for CF recommendation. Graph-based analysis is gaining momentum in recent years due to their ubiquity in real-world problems such as modeling user preference for item recommendation as well as social graphs in social media (SM) networks, co-occurrence graph in linguistics, citation graph in research, knowledge graph and so forth. The central idea of network embedding is to represent each entity in a graph with a lower-dimensional latent representation to facilitate tasks within the network and prediction over entities. For example, such latent representation makes it possible to perform prediction for supervised tasks, while the distance between node embedding vectors can serve as a useful measure in unsupervised tasks. GraphGAN can be viewed as a graph-based representation of IRGAN, where queries/items are \textit{nodes} of the graph. For a given node $v_c$, the objective of $\mathcal{G}$ is to learn the ground-truth connectivity distribution over vertices $p_{true}(v|v_c)$, whereas $\mathcal{D}$ aims to discern whether or not a connectivity should reside between vertex pairs $(v,v_c)$. GraphGan furthermore proposes the graph softmax as $\mathcal{G}$ ---instead of traditional softmax--- which appears to boost the computational efficiency of training (graph sampling and embedding learning) performed by  $\mathcal{G}$.\\
\setlength{\parindent}{15pt} [\textbf{GAN-HNBR}] From an application perspective, GAN-based graph representations have also been applied in more niche domains of RS, including personalized citation recommendation. The goal is to recommend research articles for citation by using a content-based and author-based representation~\cite{DBLP:conf/ismis/ZhangYCD18} or learning heterogeneous bibliographic network representation (HBNR). \citet{DBLP:conf/aaai/CaiHY18} propose GAN-HNBR ---a GAN-based citation recommendation model--- that can learn the optimal representation of a bibliographic network consisting of heterogeneous vertex content features such as papers and authors into a common shared latent space and provide personalized citation recommendation.\\
\setlength{\parindent}{15pt} [\textbf{CFGAN}] CFGAN has been introduced in~\cite{DBLP:conf/cikm/ChaeKKL18} to address a problem with \textit{discrete items} in IRGAN, where $\mathcal{G}$ produces at each iteration a single item index, which is a discrete entity in nature. This is different from the original GAN in the CV domain in which the output of $\mathcal{G}$ is an image (i.e., a vector). The generation of discrete item indices by $\mathcal{G}$ results in a poor sampling of items from the pool of available alternatives (i.e., samples identical to ground-truth) deteriorating the performance of $\mathcal{G}$ and $\mathcal{D}$ ---instated of improvement--- during the min-max training iteration. CFGAN introduces \textit{vector-wise
training} in which $\mathcal{G}$ generates continuous-valued vectors to avoid misleading $\mathcal{D}$, which in turn improves the performance of both $\mathcal{G}$ and $\mathcal{D}$. The authors show the improvement of CFGAN over IRGAN and GraphGAN baselines. As an example, with regards to P@20 on the Ciao dataset, the improvement is $100\%$  for CFGAN vs. IRGAN (0.45 v.s. 0.23) and $160\%$ for CFGAN vs. GraphGAN (0.45 v.s. 0.17), which turns to be a significant improvement of the recommendation accuracy.\\
\setlength{\parindent}{15pt} [\textbf{Chae et al.}] \citet{8525831} propose an auto-encoder-based GAN, in which an auto-encoder (AE) is used as $\mathcal{G}$ to model the underlying distribution of user preferences over items. The primary motivation behind this work is that conventional MF-based approaches are linear. Instead, the proposed system can generate non-linear latent factor models and uncover more complex relationships in the underlying user-item interaction matrix.\\
\setlength{\parindent}{15pt} [\textbf{VAE}] An adversarial variational auto-encoder (VAE) is adopted by~\citet{8663730}, where the authors propose the usage of a GAN to regularize the VAE by imposing an arbitrary prior to the latent representation (based on implicit feedback). Similar works can be found in~\cite{DBLP:conf/icde/TongLZSC19, DBLP:conf/cikm/LeeSM17}, which exploits a VAE to enhance the robustness of adversarial examples. The authors furthermore present the Wasserstein distance with gradient penalty.\\
\updated{
\setlength{\parindent}{15pt} [\textbf{GAN-VAE-CF}] \citet{DBLP:conf/cikm/KrishnanSSS18} design an adversarial generative network to learn inter-item interactions used to generate informative \textit{niche} negative samples paired with the positive ones to reduce the popularity bias, and  promote recommendation of long-tail products.}\\
\setlength{\parindent}{15pt} [\textbf{CALF]} Other issues of IRGAN, such as sparsity causing gradient vanishing and update instability and discrete value preventing a training to optimize using gradient descent, are addressed in~\cite{DBLP:conf/flairs/CostaD19}. The proposed solution is named convolutional adversarial latent factor model (CALF), which employs a CNN to learn correlations between embeddings and Rao-Blackwell sampling to deal with discrete values optimizing CALF.\\
\setlength{\parindent}{15pt} [\textbf{PD-GAN]} The authors of~\cite{DBLP:conf/ijcai/WuLMZZG19} propose a solution to improve diversity of CF-based recommendation with GAN based on personalized diversification.\\
\setlength{\parindent}{15pt} [\textbf{LambdaGAN]} In~\cite{DBLP:conf/ijcnn/WangZC019}, the authors propose LambdaGAN ---a GAN model with a lambda ranking strategy--- that improves the recommendation performance in a pairwise ranking setting by proposing lambda rank~\cite{DBLP:conf/cikm/YuanGJCYZ16} function into the adversarial learning of the proposed GAN-based CF framework.\\
\setlength{\parindent}{15pt} [\textbf{VAEGAN]} A variant of VAE is introduced in~\cite{DBLP:conf/ijcai/YuZCX19} to address the limited expressiveness of the inference model and latent features, which reduces the generalization performance of the model. The proposed solution, named adversarial variational autoencoder GAN (VAEGAN), is a more expressive, and flexible model that better approximates the posterior distribution by combining VAEs and GAN. This work is one of the first work to propose the application of adversarial variational Bayes (AVB)~\cite{DBLP:conf/icml/MeschederNG17} to perform the adversarial training.

\subsubsection{Context-aware Recommendation}\label{subsec:GAN_CA_rec}

Although long-term preference modeling has proven to be effective in several domains~\cite{DBLP:conf/kdd/BerkovskyF15}, recent research indicates that users' preferences are highly variable based on the user's context, e.g., time, location, and mood~\cite{DBLP:conf/recsys/KompanKB17}. Context provides the background of \textit{user objective} for using the system and can be exploited to generate more relevant recommendations.

\noindent \textbf{Temporal-aware Recommendation.}
In real applications, users' preferences change over time, and modeling such \textit{temporal evolution} is needed for effective recommendation. While long-term preferences of users change slowly, their \textit{short-term preferences} can be seen as more dynamic and changing more rapidly. Predicting short-term user preference has been recently studied in the context of \textit{session-based} and \textit{sequential recommendations}. A temporal extension of SVD++ towards the modeling of temporal dynamic,  named TimeSVD++, has been proposed in~\cite{DBLP:journals/cacm/Koren10}. It has also been reported that the structure of time-aware inputs (e.g., click-logs, session) are effectively modeled by a recurrent neural network (RNN).
For instance, \citet{DBLP:journals/corr/HidasiKBT15} proposed to model the sequential user clicks to output session-based recommendation with a GRU-gated recurrent unit; while \citet{DBLP:conf/wsdm/WuABSJ17} proposed to integrate an LSTM model, to capture both the user and the item temporal evolution, and MF to model stationary preferences. 
\updated{
Both LSTM and GRU are variants of RNNs deployed to reduce the gradient vanishing problem by including a mechanism to enable the model to learn the historical evolution of users' behavior. In particular, GRUs are generally preferred over LSTMs since they have to learn less model parameters and, consequently, require less memory than LSTMs~\cite{DBLP:conf/icml/JozefowiczZS15}. 
}
Inspired by the accuracy improvements of IRGAN, GAN-based models have been combined in temporal frameworks to boost the recommendation performance in sequence-aware recommendation tasks. 

\setlength{\parindent}{15pt} [\textbf{RecGAN}] In~\cite{DBLP:conf/recsys/BharadhwajPL18}, the authors propose to incorporate in a single framework both the temporal modeling capabilities of RNN and the latent feature modeling power of the \textit{min-max  game}. The proposed framework, named RecGAN, implements both the generator and the discriminator with the Gated Recurrent Unit (GRU)~\cite{DBLP:conf/emnlp/ChoMGBBSB14}, in order to make $\mathcal{G}$ capable of predicting a sequence of relevant items based on the dynamic evolution of user's preferences.\\
\setlength{\parindent}{15pt} [\textbf{PLASTIC \& LSIC}] Differently from RecGAN that implements only an RNN cell to capture the dynamic evolution of the user's behavior, \citet{DBLP:conf/ijcai/ZhaoWYGYC18, 8643032} propose to combine MF and RNN in an adversarial recommendation framework to model respectively long and short-term user-item associations. The proposed framework, named PLASTIC, adopts MF and LSTM cells into $\mathcal{G}$ to account for the varying aspect of both users and items,  while a two-input Siamese network ---built manually by using a MF and RNN--- as $\mathcal{D}$ encodes both the \textit{long-term} and \textit{session-based information} in the pair-wise scenario.\\
\setlength{\parindent}{15pt} [\textbf{NMRN-GAN}] Recent studies have endorsed that adversarially created close-to-observed negative samples  are capable of improving the user and item representation. \citet{DBLP:conf/kdd/WangYHLWH18} introduce \textit{GAN-based negative sampling} for streaming recommendation. Instead of using a random sampling strategy, which is static and hardly contributes towards the training of the recommender model, adversarially generated negative samples result more informative. NMRN-GAN uses a key-value memory network~\cite{DBLP:conf/www/ZhangSKY17} to keep the model's long-term and short-term memory combined with a GAN-based negative sampling strategy to create more instructive negative samples thus improving the training effectiveness and the quality of the recommendation model.\\
\setlength{\parindent}{15pt} [\textbf{GAN-CQDN}] A GAN-based solution has been proposed in~\cite{DBLP:conf/icml/Chen0LJQS19} for sequence-aware recommendation in conjunction with reinforcement learning (RL). The main aim here is that of modeling the dynamic of user's status and long-term performance. The authors propose GAN-CQDN, an RL-based recommender system that exploits GAN to model user behavior dynamics and learn her reward function. The advantages of using GAN is that it improves the representation of the user profile a well as the reward function according to the learned user profile, and it accommodates online changes for new users. \\
\updated{
\setlength{\parindent}{15pt} [\textbf{MFGAN}] \citet{DBLP:conf/sigir/RenLLZWDW20} implement the generator as a Transformer-based network to predict, for each user, the next relevant item whose importance will be judged from multiple factor-specific discriminators. The discriminators are a set of Transformed-based binary classifiers that measure the recommendation relevance with respect to additional information (e.g., item popularity, semantic information, and price). Similar to MFGAN~\cite{DBLP:conf/sigir/RenLLZWDW20}, \citet{DBLP:conf/ijcai/ZhaoSZXB20} propose an Adversarial Oracular Seq2seq learning for sequential Recommendation (AOS4Rec) framework to enhance the recommendation performance of Transformer-based, or RNN-based, next-item recommenders.
}\\
\noindent \textbf{Geographical-aware Recommendation.}  Another relevant application of contextual information is point-of-interest (POI) recommendation. In this field, many approaches have been proposed over the year especially after the mobile revolution. Location-based social networks (LBSNs) have attracted millions of users to share rich information, such  as experiences and  tips.  Point-of-Interest  (POI) recommender systems play an important role in LBSNs since they can help users explore attractive locations as well as help social network service providers design location-aware advertisements for Point-of-Interest.\\
\setlength{\parindent}{15pt} [\textbf{Geo-ALM}] In~\cite{DBLP:conf/ijcai/0061W0Y19}, the authors propose Geo-ALM, a GAN-based POI recommender that integrates geographical features (POI and region features) with a GAN to achieve (better) POI recommendation. In the proposed system, $\mathcal{G}$ improves the random negative sampling approach in the pairwise POI recommendation scenario that leads to better representation of user and items and enhances recommendation quality with respect to state-of-the-art models.\\
\setlength{\parindent}{15pt} [\textbf{APOIR}] Inspired by the advances of POI recommendation performance under GAN-based framework, \citet{DBLP:conf/www/0002YZTZW19} propose adversarial point-of-interest recommendation (APOIR) to learn user-latent representations in a generative manner. The main novelty of the proposed framework is the use of POIs' geographical features and the users' social relations into the reward function used to optimize the $\mathcal{G}$. The reward function acts like a contextual-aware regularizer of $\mathcal{G}$, that is the component of APOIR in the proposed POI recommendation model.

\subsubsection{Cross-domain Recommendation}\label{subsec:GAN_CR_rec}

Recommender models are usually designed to compute recommendations for items belonging to a single domain. Items belonging to a specific domain share characteristics and attributes, which are intrinsically similar, and domain-specific recommendation models allow the designer to study these characteristics individually. However, \textit{single-domain} recommendation faces numerous challenges. The first challenge refers to the well-known \textit{cold-start} problem, when insufficient interactions exist in the considered domain. Second, users' interests and needs span across different application areas and large e-commerce sites, like Amazon or eBay, store users' preference scores related to products/services of various domains ---from books and products to online movies and music. As companies strive to increase the diversity of products or services to users, cross-domain recommendation can help such companies to increase sales productivity by offering personalized cross-selling or bundle recommendations for items from multiple domains~\cite{cantador2015cross}. The third aspect is a novel research idea related to discovering relationships between items (e.g., images) of two different domains. For example, can a machine achieve a human-level understanding to recommend a fashion item consistent with user taste/style in another domain such as media or visual scenery? 

\setlength{\parindent}{15pt} [\textbf{FR-DiscoGAN}]
In~\cite{DBLP:conf/bigcomp/JoJCJ19}, the authors propose a cross-domain GAN to generate fashion designs from the sceneries. In the proposed hypothetical scenario, the user can specify via a query her POI to visit (e.g., mountain, beach) together with keywords describing a season (i.e., spring, summer, fall, and winter). The core idea is to automatically generate fashion items (e.g., clothes, handbags, and shoes) whose useful features (i.e., style) match the natural scenery specified by the user. For instance, the system can recommend a collection of fashion items that look cool/bright for visiting a beach in summer, even though the actual preference of the user is black-style clothes. The role of GAN is to learn associations between scenery and fashion images. In the field of ML and CV, the problem is termed as \dquotes{style transfer} or \dquotes{image to image translation} problem~\cite{DBLP:conf/cvpr/GatysEB16}. \\
\setlength{\parindent}{15pt} [\textbf{VAE-GAN-CC}]
An effective cross-domain recommendation system relies on \textit{capturing both similarities and differences} among features of domains and exploiting them for improving recommendation quality in multiple domains. Single-domain algorithms have difficulty in uncovering the specific characteristics of each domain. To solve this problem, some approaches extract latent features of the domains by a separate network~\cite{DBLP:conf/www/LianZXS17,DBLP:journals/tmm/MinBXH15}. Although these approaches might be successful in capturing characteristic features of each domain, they do not establish the similarity between features of multiple domains. To extract both homogeneous and divergent features in multiple domains, \citet{DBLP:journals/corr/abs-1812-06229} propose a generic cross-domain recommendation system that takes as input the user interaction history (click vector) in each domain,  maps the vectors to a shared latent space using two AEs and then uses $\mathcal{G}$ to remap the underlying latent representation to click vectors. The main novelty of this work lies in building/linking shared latent space between domains, which in turn facilitates \textit{domain-to-domain} translation. In particular, the former is realized by enforcing a weight-sharing constraint related to variational auto-encoders, i.e., the encoder-generator pair $\{\mathcal{E}_A, \mathcal{G}_A\}$ and $\{\mathcal{E}_B, \mathcal{G}_B\}$ and using cycle-consistency (CC) as a weight-sharing constraint. Finally, two separate adversarial discriminators are employed to determine whether the translated vectors are realistic. The final system is called VAE-GAN-CC network, which extends the unsupervised image-to-image translation network in the CV domain~\cite{DBLP:conf/nips/LiuBK17} for RS applications and is thus named domain-to-domain translation model (D2D-TM).\\
\setlength{\parindent}{15pt} [\textbf{DASO}]
Inspired by the efficacy of \textit{adversarial negative sampling} techniques proposed by~\citet{DBLP:conf/kdd/WangYHLWH18}, \citet{DBLP:conf/ijcai/FanD0WTL19} address the limitation of typical negative sampling in the \textit{social recommendation} domain in transferring users' information from social domain to item domain. The proposed Deep Adversarial SOcial recommendation (DASO) system, harnesses the power of adversarial learning to dynamically generate difficult negative samples for \textit{user-item} and \textit{user-user pairs}, to guide the network to learn better user and item representations. \\
\updated{
\setlength{\parindent}{15pt} [\textbf{Asr}] \citet{DBLP:conf/cikm/KrishnanCTS19} propose an adversarial social regularization (Asr) framework to improve the item recommendation performance by integrating contextual information, i.e., users' social connections, within a GAN-based approach. Furthermore, the proposed framework is agnostic to the recommender models guaranteeing applicability in several settings and domains. For instance, the authors demonstrate the framework's efficacy in within a large set of models (e.g., VAE).}\\
\setlength{\parindent}{15pt} [\textbf{CnGAN}]
\citet{DBLP:conf/www/PereraZ19}, propose GAN for cross-network (CnGAN) to address one of the significant shortcomings of cross-network recommendation concerning \textit{non-overlapping users}  missing preference scores. These users exist in the source domain but not in the target domain, and thus, their preferences about items in the target domain are not available. In the proposed work, $\mathcal{G}$ learns the mapping of user preferences from \textit{target to source} and generate more \textit{informative preferences} on the source domain. $\mathcal{D}$ uses the synthetically generated preferences (generated from $\mathcal{G}$) to provide recommendations for users who only have interactions on the target network (not overlapped users). The authors also propose two novel loss functions ---a content-wise and a user-wise loss function--- to guide the min-max training process better. The authors validate the effectiveness of the system against state-of-the-art models both in terms of accuracy and beyond-accuracy measures (novelty, diversity).

\subsubsection{Fashion Recommendation}
\label{subsec:GAN_FA_rec}
Most conventional RS are not suitable for application in the fashion domain due to unique characteristics hidden in this domain. For instance, people do not follow the crowd blindly when buying clothes or do not buy a fashion item twice~\cite{DBLP:conf/waim/ShaWZFZY16}. 
Another aspect is related to the notion of  \textit{complementary} relationship for recommending a personalized fashion outfit. It is natural for humans to establish a sense of relationship between products based on their visual appearance. 
Recently, GAN-based models have shown promising performance for outfit recommendation, being able to compete with state-of-the-art fashion recommendation models in the field, such as Siamese-base networks~\cite{DBLP:journals/mms/GaoLWZ19}. Finally, another new application of GANs is related to exploiting the \textit{generative power of GANs} to synthesize real-looking fashion clothes. This aspect can inspire the aesthetic appeal/curiosity of costumer and designers and motivates them to explore the space of
potential fashion styles.

\setlength{\parindent}{15pt} [\textbf{CRAFT}] \citet{DBLP:conf/eccv/HuynhCTA18} address the problem of recommending complementary fashion items based on visual features by using an adversarial process that resembles GAN and uses a conditional feature transformer as $\mathcal{G}$ and a discriminator $\mathcal{D}$. One main distinction between this work and the prior literature is that the $\langle$input, output$\rangle$ pair for $\mathcal{G}$ are both features (here features are extracted using pre-trained CNNs~\cite{DBLP:conf/aaai/SzegedyIVA17}), instead of $\langle$image, image$\rangle$ or hybrid types such as $\langle$image, features$\rangle$ explored in numerous previous works~\cite{DBLP:conf/iccv/ZhuFULL17,DBLP:conf/cvpr/VolpiMSM18}. This would allow the network to learn the relationship between items directly on the feature space, spanned by the features extracted. The proposed system is named complementary recommendation using adversarial feature transform (CRAFT) since in the model, $\mathcal{G}$ acts like a feature transformer that ---for a given query product image $q$--- maps the source feature $s_{q}$ into a complementary target feature $\hat{t}_{q}$ by playing a min-max game with $\mathcal{D}$ with the aim to classify fake/real features. For training, the system relies on learning the co-occurrence of item pairs in real images. In summary, the proposed method does not generate new images; instead it learns how to generate features of the complementary items conditioned on the query item. 

\setlength{\parindent}{15pt} [\textbf{DVBPR}] Deep visual Bayesian personalized ranking (DVBPR)~\cite{DBLP:conf/icdm/KangFWM17} is presumably one of the first works that exploit the \textit{visual generative power of the GAN} in the fashion recommendation domain. It aims at generating clothing images based on user preferences. Given a user and a fashion item category (e.g., tops, t-shirts, and shoes), the proposed system generates new images ---i.e., clothing items--- that are consistent with the user's preferences. The contributions of this work are two-fold: first, it builds and end-to-end learning framework based on the Siamese-CNN framework. Instead of using the features extracted in advance, it constructs an end-to-end system that turns out to improve the visual representation of images. Second, it uses a GAN-based framework to generate images that are consistent with the user's taste. Iteratively, $\mathcal{G}$ learns to generate a product image integrating a \textit{user preference maximization objective}, while $\mathcal{D}$ tries to distinguish crafted images from real ones. Generated images are quantitatively compared with real images using the preference score (mean objective value), inception score~\cite{DBLP:conf/nips/SalimansGZCRCC16}, and opposite SSIM~\cite{DBLP:conf/icml/OdenaOS17}. This comparison shows an improvement in preference prediction in comparison with non-GAN based images. At the same time, the qualitative comparison demonstrates that the generated images are realistic and plausible, yet they are quite different from any images in the original dataset ---they have standard shape and color profiles, but quite different styles.

\setlength{\parindent}{15pt} [\textbf{MrCGAN}] \citet{DBLP:conf/aaai/ShihCLS18} propose a compatibility learning framework that allows the user to visually explore candidate \textit{compatible prototypes} (e.g., a white T-shirt and a pair of blue-jeans). The system uses metric-regularized conditional GAN (MrCGAN) to pursue the item generation task. It takes as the input a projected prototype (i.e., the transformation of a query image in the latent "Compatibility Space"). It produces as the output a synthesized image of a compatible item (the authors consider a compatibility notion based on the complementary of the query item across different catalog categories). Similar to the evaluation protocol in~\cite{DBLP:conf/eccv/HuynhCTA18}, the authors conduct online user surveys to evaluate whether their model could produce images that are perceived as compatible. The results show that MrCGAN can generate compatible and realistic images under compatibility learning setting compared to baselines.

\setlength{\parindent}{15pt} [\textbf{Yang et al.} \& \textbf{$c^+$GAN}] \citet{Yang2018FromRT} address the same problem settings of MrCGAN~\cite{DBLP:conf/aaai/ShihCLS18} by proposing a fashion clothing framework composed of two parts: a clothing recommendation model based on BPR combined with visual features and a clothing complementary item generation based GAN. Notably, the generation component takes in input a piece of clothing recommended in the recommendation model and generates clothing images of other categories (i.e., top, bottom, or shoes) to build up a set of complementary items. The authors follow a similar qualitative and quantitative evaluation procedure as DVBPR~\cite{DBLP:conf/icdm/KangFWM17} and further propose a \textit{compatibility index} to measure the compatibility of the generated set of complementary items. A similar approach has also been proposed in $c^+$GAN~\cite{DBLP:journals/corr/abs-1906-05596}, to generate bottom fashion item paired with a given top item.

\section{Summary and Future Directions}\label{sec:conclusion}
In this paper, we have surveyed a wide variety of tasks in which adversarial machine learning (AML) is important to attack/defense a recommendation model as well as improve the generalization performance of the model itself. This broad range of applications can be categorized into two ---objective-wise distinct--- technologies: (i) AML for improving security (cf. Section~\ref{sec:security}) and, (ii) AML used in generative adversarial networks (GANs) exploited for numerous tasks such as better CF recommendation, context-aware recommendation, cross-domain system, or visually-aware fashion item/outfit recommendation (cf. Section~\ref{sec:GAN}). The common point of both technologies is the joint min-max optimization used for training models, in which two competing players play a zero-sum differential game until they reach an equilibrium. To the best of our knowledge, this is the first work that sums up the advances of AML application in recommendation settings and proposes a clear taxonomy to classify such applications.

We put forward what is better to invest in AML-RS research and introduce the following open research directions:

\noindent\emph{\underline{Bridging the gap between attack/defense models in the ML/CV and RS domain.}} 
As the prior literature of AML for security emerged in the field of machine learning (ML) and computer vision (CV), there remains a large gap between advances made in those fields and that in RS. Consider the questions: \dquotes{Attacks for images are designed to be human-imperceptible or inconspicuous (i.e., may be visible but not suspicious).  How can we capture these notions for designing attacks in RS?}; furthermore, \dquotes{Images are continuous-valued data while a user profile is a discrete data. Modifying users' profiles completely changes the semantic of their behaviors. What is the best approach to treat these nuances in RS attack designs?} \\
\emph{\underline{Choice of recommendation models.}} Modern recommendation models exploit a
wealth of side-information beyond the user-item matrix such as social-connections, multimedia content, semantic data, among others. However, most of the attacks against recommendation systems are designed and validated against CF systems. Investigating the impact of adversarial attacks against these ---heterogeneous in nature--- data types remains as an open highly interesting challenge, e.g, consider adversarial attacks against music, image, and video recommendation models leveraging multimedia content. In this regard, we also recognize attack against state-of-the-art deep and graph-based models, another highly-valued research direction.\\
\emph{\underline{Definition of attack threat model.}} The research in RS community misses a common evaluation approach for attacking/defending scenarios such as the one introduced by Carlini at el.~\cite{DBLP:journals/corr/abs-1902-06705}. For instance, it is important to define a common attacker threat model to establish in advance the attacker knowledge and capabilities to make the attack (or defense) reproducible and comparable with novel proposals.\\
\emph{\underline{Move the attention towards beyond accuracy goal in recommendation.}}
According to our survey, most of the identified research works focus on accuracy metrics such as HR and nDCG. Consider the question: \dquotes{What is the impact of adversarial attacks and defenses in other evaluation objectives of RS, for instance, diversity, novelty, and fairness of recommendations}. The impact on these metrics could be, in principle, the main objective of a new breed of attack strategies aiming at compromise the diversity/novelty of results.\\
\emph{\underline{Scalability and stability of learning.}} We identify that there exists the need to further explore the stability learning problems in the discrete item sampling strategy to train the generator. This has been already identified as a big problem when GAN-based RS are applied in real scenarios with huge catalogues. A point of study may be that of novel GAN models proposed in computer vision (e.g., WGAN~\cite{DBLP:journals/corr/ArjovskyCB17}, LSGAN~\cite{DBLP:conf/iccv/MaoLXLWS17}, and BEGAN~\cite{DBLP:journals/corr/BerthelotSM17}).\\
\emph{\underline{Users preferences learning with GANs.}} An interesting and already established application of AML-RS is to exploit the generative power of GANs to produce more plausible user-rating profiles that can be used to improve recommendations in the cold-user scenario or improve the prediction performance in warm-start settings. We consider such applications extremely interesting, and we motivate further research in this direction to resolve the well-known cold-start obstacles in recommendation settings.





\bibliographystyle{ACM-Reference-Format}
\bibliography{refs_final}

\end{document}